\documentclass[10pt,showpacs,showkeys,nofootinbib]{revtex4-1}
\usepackage[T1]{fontenc}
\usepackage{graphicx}
\usepackage{dcolumn}
\usepackage{bm}
\usepackage{amsmath} 
\usepackage[english]{babel}
\usepackage{mathrsfs}
\usepackage{amsfonts}
\usepackage{amssymb}
\usepackage[latin1]{inputenc}
\usepackage{hyperref}
\usepackage{longtable}
\usepackage{dcolumn}
\usepackage{appendix}
\usepackage{multirow}
\usepackage{natbib}  
\usepackage{color}

\newcommand{\mytilde}{\raise.17ex\hbox{$\scriptstyle\mathtt{\sim}$}}
\newcommand{\barr}{\begin{eqnarray}}
\newcommand{\earr}{\end{eqnarray}}
\newcommand{\bea}{\begin{eqnarray*}}
\newcommand{\eea}{\end{eqnarray*}}
\newcommand{\beq}{\begin{equation}}
\newcommand{\eeq}{\end{equation}}
\newcommand{\tc}{\eta_{\vec{k}}^c}
\newcommand{\nn}{\nonumber \\}
\newcommand{\zk}{|z_k|}
\newcommand{\mH}{\mathcal{H}}

\newcommand{\dphi}{\delta \phi}
\newcommand{\vx}{\vec{x}}

\newcommand{\bra}{\langle}
\newcommand{\ket}{\rangle}
\newcommand{\nk}{\vec{k}}
\newcommand{\ann}{\hat{a}} 
\newcommand{\cre}{\hat{a}^{\dagger}}    

\newcommand{\mR}{\mathcal{R}}

\newcommand{\Mpc}{\mathrm{Mpc}}

\newcommand{\A}{$\mathcal{A}$~}
\newcommand{\B}{$\mathcal{B}$~}
\newcommand{\RI}{\textrm{R,I}} 
\newcommand{\yk}{y_{\vec{k}}}
\newcommand{\pk}{\pi_{\vec{k}}}

\begin{document}

 \title{Constraining quantum collapse inflationary models with current data: The semiclassical approach}

\author{ Mar\'{\i}a P\'{\i}a Piccirilli$^1$\footnote{E-mail: mpp@fcaglp.unlp.edu.ar }}
\author{Gabriel Le\'{o}n$^1$}
\author{Susana J. Landau$^2$}
\author{Micol Benetti$^3$$^4$$^5$}
\author{Daniel Sudarsky$^6$}

\affiliation{$^1$Grupo de Astrof\'{\i}sica, Relatividad y
  Cosmolog\'{\i}a, Facultad de Ciencias Astron\'{o}micas y
  Geof\'{\i}sicas, Universidad Nacional de La Plata, Paseo del Bosque
  S/N 1900 La Plata, Pcia de Buenos Aires, Argentina.}

\affiliation{$^2$Departamento de F\'isica, Facultad de Ciencias
  Exactas y Naturales, Universidad de Buenos Aires and IFIBA, CONICET,
  Ciudad Universitaria - PabI, Buenos Aires 1428, Argentina}

\affiliation{$^3$Observat\'{o}rio Nacional, Rua General Jos\'{e} Cristino 77, 20921-400, Rio de Janeiro, RJ, Brazil}
\affiliation{$^4$University of Naples Federico II
Physics Department "Ettore Pancini"
Monte Sant'Angelo Campus
Via Cinthia 21, I-80126 Naples, Italy}
\affiliation{$^5$Istituto Nazionale di Fisica Nucleare (INFN), Sez. Napoli,
Monte Sant'Angelo Campus
Via Cinthia 9, I-80126 Naples, Italy}

\affiliation{$^6$Instituto de Ciencias Nucleares, Universidad
  Nacional Aut\'{o}noma de M\'{e}xico, A.P. 70-543, M\'{e}xico
  D.F. 04510, M\'{e}xico.}

\date{Dec 2018}

\begin{abstract}
The hypothesis of the self-induced collapse of the inflaton wave
  function was introduced as a candidate for the physical process
  res\-pon\-si\-ble for the emergence of inhomogeneity and anisotropy
  at all scales.  In particular, we consider different proposal for
  the precise form of the dynamics of the inflaton wave function: i)
  the GRW-type \emph{collapse schemes} proposals based on spontaneous
  individual collapses which generate non-vanishing expectation values
  of various physical quantities taken as \emph{ansatz} modifications
  of the standard inflationary scenario; ii) the proposal based on a
  Continuous Spontaneous Localization (CSL) type modification of the
  Schr\"odinger evolution of the inflaton wave function, based on a
  natural choice of collapse operator. We perform a systematic
  analysis within the semi-classical gravity approximation, of the
  standing of those models considering a full quasi-de sitter
  expansion scenario. We note that the predictions for the Cosmic
  Microwave Background (CMB) temperature and polarization spectrum
  differ slightly from those of the standard cosmological model. We
  also analyse these proposals with a Bayesian model comparison using
  recent CMB and Baryonic Acoustic Oscillations (BAO)
  data. 
  Our results show a moderate preference of the joint CMB and BAO data
  for one of the studied \emph{collapse schemes} model over the
  $\Lambda$CDM one, while there is no preference when only CMB data
  are considered. Additionally, analysis using CMB data provide the
  same Bayesian evidence for both the CSL and standard models,
  i.e. the data have not preference between the simplicity of the LCDM
  model and the complexity of the collapse scenario.
\end{abstract}



\maketitle
\section{Introduction}
\label{intro}

The assumption of an inflationary period at the very early stages of
the universe's history is usually considered part of the standard
cosmological model
\cite{starobinsky1980,guth1980,linde1981,albrecht1982}, and the
physics of such period is viewed as providing an account for the
observed cosmic structure
\cite{mukhanov1981,mukhanov1982,starobinsky1982,guth1982,hawking1982,bardeen1983}. According
to this picture, during the inflationary era, the evolution of the
universe is described by a Friedmann-Robertson-Walker (FRW) background
cosmology with an accelerated expansion. In the simplest inflationary
model, the expansion is driven by the potential of a single scalar
field: the inflaton.  Additionally, the quantum fluctuations of the
inflaton are characterized by a simple vacuum state that is exactly
symmetric, being the symmetry the homogeneity and isotropy of the
quantum state.  However, when considering the standard inflationary
scenario more carefully an important issue arises, namely, the
transition from a perfect symmetric state in the early universe to the
present non-symmetric state of the current universe, which cannot be
attributed to the quantum unitary evolution.  This shortcoming of the
inflationary scenario has been extensively discussed
\cite{PSS06,Shortcomings,LLS13} and a proposal to deal with has been
developed
\cite{PSS06,Shortcomings,US08,Leon10,DT11,Leon11,LSS12,LLS13,LLP15,CPS13}.
The most important ingredient of this proposal is to introduce
\emph{the self-induced collapse hypothesis}: an internally induced
collapse of the inflaton wave function as the physical mechanism
responsible for the emergence of inhomogeneities and anisotropies at
each particular length scale. It must be emphasized that we are not
calling into question the standard inflation/$\Lambda$CDM
paradigm. Our proposal simply incorporates to the inflationary model a
physical process capable of turning the homogeneity and isotropy of
the vacuum state into actual inhomogeneities and anisotropies.

Our formulation of the \emph{collapse proposal} assumes that at the
early inflationary stage during the cosmic evolution, there was a
spontaneous ``jump'', or a continuous series of infinitesimal
``jumps'', of the quantum state associated to a particular mode of the
quantum field. That is very similar to what is usually taken to
characterize the measurement process of Quantum Mechanics, which
results in the quantum mechanical collapse of the wave
function. However, in this approach, there is no external measuring
device or observer that is responsible for triggering such
collapse. The question one faces then is to explain the physical
mechanism responsible for such spontaneous collapse. Various authors
have proposed that the collapse of the wave function is a physical
process induced by unknown aspects of quantum gravity
\cite{penrose1996,diosi1987,diosi1989,ghirardi1985}. On the other
hand, various proposals based on an objective dynamical reduction of
the wave function have been developed in different contexts than the
cosmological one
\cite{ghirardi1985,pearle1989,bassi2003,bassi2013}. The aim of those
proposals is to provide a solution to the quantum measurement problem,
which in the particular case of cosmology is exacerbated by the
absence of a well defined notion of observers, measurement devices,
that might play a special role in the early universe.

It is also important to mention that the conceptual issue we are
discussing is sometimes referred in the literature as the
quantum-to-classical transition of the primordial perturbations
{\cite{grishchuk,grishchuk1992,polarski,Lesgourgues1996,Grishchuk1997,kiefer}}.\footnote{In
  fact, we find that characterizing the problem as the
  ``quantum-to-classical'' transition is somewhat misleading. Our
  posture is that there are no classical or quantum regimes. The
  fundamental description is always quantum mechanical. However, in
  some physical systems, there exist certain conditions that allow us
  to describe specific quantities, to a sufficient accuracy, by their
  expectation values which are then identified with their classical
  counterparts.}  We note in this regard, some authors argue that
decoherence
\cite{grishchuk,polarski,Lesgourgues1996,kiefer,Kiefer2006,Egusquiza1997,Burgess2006}
and/or the squeezing nature of the evolved vacuum state of the
inflaton \cite{grishchuk1992,polarski,albrecht1992} provides a
complete resolution of the problem. Nevertheless, as discussed in
detail in \cite{Shortcomings}, we do not endorse to such claims mainly
because the squeezed nature of a quantum state can not be taken as an
indication that the system has become classical, nor that the
symmetries of the quantum state have disappeared. In fact, one can
always find a new set of operators in which the evolved (squeezed)
state will look like a ``standard vacuum.'' For instance, if we
consider the simple quantum harmonic oscillator, we can write the
usual creator and annihilation operators as $\hat a =
(1/\sqrt{2})(e^{s_0} \hat x + i e^{-s_0} \hat p)$, $\hat a^\dagger =
(1/\sqrt{2})(e^{s_0} \hat x - i e^{-s_0} \hat p)$ with $e^{s_0} =
\sqrt{m\omega}$ and usual commutator relations. We can now define
(without changing the system or its Hamiltonian) for arbitrary values
of $s$, new operators $\hat a_s = (1/\sqrt{2})(e^{s} \hat x + i e^{-s}
\hat p)$, $\hat a_s^\dagger = (1/\sqrt{2})(e^{s} \hat x - i e^{-s}
\hat p)$ which are related to the original creation and annihilation
operators through a ``Lorentzian rotation'' and obey $[\hat a_s, \hat
a_s^\dagger]=1$. However following the reasoning of e.g. \cite{kiefer}
we would have that in the limit when $s \to \infty$, the fact that
$\hat x$ and $\hat p$ do not commute becomes irrelevant and therefore
we are in an essentially classical situation, where for each value of
$\hat a_s$ there is a corresponding value of $\hat a_s^\dagger$. The
former argument would imply that, through the simple act of choosing
to express things in terms of suitable variables, we can change the
nature of a purely quantum mechanical system into something which is
essentially classical. The direct connection with the inflationary
scenario is evidenced when nothing that the former construction is
just what is used to define squeezed states. Additionally, that
construction is precisely what characterizes the Bogoliubov connection
between the creation annihilation operators that can be taken as
natural (in the sense of leading to a simple form for the
instantaneous Hamiltonian) in the early times and the corresponding
ones for late times.  Another approach to address the so called
classicalization of the primordial perturbations, point to the Everett
``many-worlds'' interpretation of quantum mechanics
\cite{nomura2011,nomura2011b,mukhanov2005}. However, it has been shown
that none of these approaches (decoherence, squeezing of the vacuum
state, many-worlds, etc.) can offer a completely satisfactory solution
to the problem at hand (see
refs.~\cite{PSS06,Shortcomings,Leon10,Leon11,LLS13} for the conceptual
and formal details of this issue). Other attempts based on non-local
hidden variable theories have been considered
\cite{valentini2008,pintoneto2011}, but it must be noted that, just as
the present one, those approaches go beyond standard quantum theory.

As we have already mentioned, once we assume the self-induced collapse
of the wave function, the following task is to precisely describe the
dynamics of such a process. For this purpose, there are two approaches
that have been developed recently: i) a purely phenomenological
approach, which is described by a general parametrization of the
quantum state after collapse; we will refer to this as \emph{collapse
  schemes} approach \cite{US08,LSS12,LLP15,BLA16} ii) the proposal
based on the use of an adaptation to cosmology of the Continuous
Spontaneous Localization (CSL) model
\cite{pearle1989,bassi2003,bassi2013} where a modification of the
Schr\"odinger equation leads naturally to the eventual collapse of the
inflaton wave function \cite{CPS13}.

In the \emph{collapse schemes} approach, we characterize the
post-collapse state by the quantum expectation values of the field and
its conjugated momentum. As a consequence of the collapse, those
expectation values change from being zero, when evaluated in the
vacuum state, to having non-vanishing value in the post-collapse
state. Each \emph{collapse scheme} leads to a particular pattern for
the post-collapse expectation values, leaving an imprint in the
primordial power spectrum.  As a consequence the predictions for the
Cosmic Microwave Background (CMB) spectrum changes with respect to the
standard inflationary models one. Therefore, it can be used to
constrain these models with recent observational data. In previous
works, the so called \emph{independent}, \emph{Newtonian} and
\emph{Wigner} schemes have been carefully analyzed
\cite{US08,LSS12,LLP15,BLA16}. In all cases, it has been obtained for
the primordial power spectrum of scalar perturbations, an expression
of the form $P(k) = A_s k^{n_s - 1 } Q(k)$ where $Q(k)$ is a function
characterizing the details of the specific hypothesis regarding the
collapse, and which, in particular takes different forms for each
collapse scheme.  It has been shown that for those collapse schemes,
if the conformal time of collapse of each mode of the field is given
by $\tc = {\mathcal{A}}/{k}$ with \A being a constant, then the
standard prediction for primordial power spectrum of the standard
inflationary accounts is recovered. Furthermore, two deviations from
the aforementioned parametrization of the time of collapse have been
proposed: i) $\tc = {\mathcal{A}}/{k} + {\mathcal{B}}/{k^2}$ and ii)
$\tc = {\mathcal{A}}/{k} + \mathcal{B}$ with $\mathcal{B}$ a
constant. Both type of deviations lead to modifications in the
temperature and polarization power spectrum:  for  models based on  i) the modifications are  most pronounced  at large angular scales (i.e. lowest multipoles),  while the effect for the models based on ii) can be seen   mainly in the smallest angular scales.

Comparison with recent CMB data, namely from the Planck collaboration
(2015), has been performed for the case i) of the collapse time. In
particular, we have shown that the \emph{Wigner scheme} scheme
provides the same Bayesian evidence as the minimal standard
cosmological model ($\Lambda$CDM),\footnote{By standard cosmological
  model ($\Lambda$CDM) we understand a specific choice of the
  cosmological parameters plus the standard inflationary model, in
  contrast with the collapse models, where the collapse hypothesis is
  assumed for inflation and the cosmological parameters remain
  unchanged} while the \emph{New\-to\-nian} scheme is weakly
disfavored with respect to the standard cosmology. On the other hand,
comparison with data from the WMAP collaboration and Sloan Digital Sky
Survey with the latter choice for the conformal time, has been
performed in ref.~\cite{LSS12}. However, in the parametrization
corresponding to case (ii), the underlying theoretical model was based
on an almost exact de Sitter background ($\dot H = $ constant $\neq
0$), which resulted in a theoretical prediction for the power spectrum
with $n_s = 1$. A more realistic theoretical analysis was carried out
in ref.~\cite{LLP15}, where a quasi-de Sitter background was
considered. In the present work, we compare the predictions of the
\emph{Newton} and \emph{Wigner collapse schemes} for the second choice
of the conformal time of collapse (ii) \cite{LLP15}, with recent CMB
data from the Planck collaboration (2015) and Baryonic Acoustic
Oscillations (BAO)
data. 

The application of the CSL collapse model, to inflation was first
analyzed in the context of semiclassical gravity with an (almost)
exact de Sitter background in Ref.  \cite{CPS13}. Furthermore the use
of CSL model in the context of inflation as treated instead in terms
of the Mukhanov-Sasaki variable (which involves the quantization of
both the inflaton filed and the metric perturbations) has been
considered in
\cite{DLS11,jmartin,tpsingh,tpsingh2,LB15,MBL16,Banerjee2016} (see
also \cite{magueijo2016}).  In this work, we extend the analysis
within the semiclassical gravity framework while considering a
quasi-de Sitter background metric. Moreover, we perform a statistical
analysis to compare the predictions obtained using the CSL approach,
with recent CMB data.

For all the models analyzed in this paper, we have performed a
Bayesian statistical analysis and a Bayesian model comparison in order
to determine, irrespective of all conceptual issues, whether the data
support such a scenario when compared to the standard one based on a
minimal $\Lambda$CDM model.

The paper is organized as follows: In section \ref{model}, we briefly
review the collapse hypothesis within the semiclassical gravity
approximation and summarize both collapse mechanisms analyzed in this
paper: the \emph{collapse scheme} approach and the one based on the
adaptation to the cosmological setting of the CSL model. In section
\ref{primordialpower}, we review the expressions for the primordial
power spectrum calculated in ref.~\cite{LLP15} for the \emph{collapse
  schemes}. Furthermore, we calculate the primordial power spectrum
for the CSL model in a quasi-de Sitter background for the first
time. In addition, we analyze the effect of the proposed treatments on
the CMB temperature angular spectrum and describe the results in terms
of an appropriate parametrization. Afterwards, in section
\ref{Method}, we introduce the computational and statistical tools,
and the data set used in our
analysis. 
In section \ref{results}, we present the results of our analysis and
the constraints on the cosmological and collapse parameters. Finally,
in section \ref{conclusions}, we summarize the main results of the
paper and present our conclusions.

\section{The model}
\label{model}

In this section, we briefly review the key aspects of inflationary
models with a self-induced collapse of the inflaton wave function. For
a detailed description of this framework, we refer the reader to
refs.~\cite{PSS06,US08,LSS12,LLP15}. Regarding notation and
conventions, we will work with signature $(-,+,+,+)$ for the metric;
primes over functions will denote derivatives with respect to the
conformal time $\eta$, and we will use units where $c=\hbar=1$ but
keep the gravitational constant $G$.

At this point it is worthwhile clarifying the general ideology behind
the manner we investigate the interface between gravitation and
quantum. Most modern research programs concerned about this question
start by postulate the full quantum gravity theory inspired by lines
of thought that lead to conjectures about what ought to be taken as
fundamental language to deal with the issue, mathematical elegance and
so forth, (for instance LQG or String theory or Causal Sets approach)
and then the program seeks to make connection with theories whose
validity in the appropriate regime is taken as well established
(e.g. General Relativity or Quantum Field Theory) or even sometimes
the quest is directly to seek connections with the empirical
accessible world.  These approaches often face the questions of how to
connect with observations, and substantial amount of work, some of it
involving ``reasonable guesses'' (but guesses nonetheless), to even
reach that point. That is, what we call the \textit{bottom--up
  approach} to physics. We approach the issue in the opposite
direction, using what we call the \textit{top--down approach}: This is
based on being agnostic regarding the nature of the fundamental
theory, and instead considering extrapolation of rather well
stablished theories (General Relativity together with Quantum Field
Theory in curved space-time) to regimes where such inquires are
expected to face very delicate issues. We then try to find simple
manners to bridge these difficulties and then study whether or not the
results are reasonable (i.e do they agree with observations?  or do
they fit together with other clues about the regimes of interest?).
This is not to be construed as a criticism to the standard
approaches. We in fact think the two are reasonable paths of inquire,
but the point is that we must recognize that each one of those faces
the most difficult problems are at different stages of the program.

As in standard slow-roll inflationary models, we consider the action
of a single scalar field, minimally coupled to gravity, with an
appropriate potential:
\begin{equation}
S[\phi,g_{ab}] = \int d^4x \sqrt{-g} \bigg[ \frac{1}{16 \pi G} R[g] 
- \frac{1}{2}\nabla_a \phi \nabla_b \phi g^{ab} - V[\phi] \bigg]. 
\label{actioncol}
\end{equation}
The background metric is described by a FRW spacetime. In conformal
coordinates, the components of the background metric are $ g_{\mu
  \nu}^{(0)} = a(\eta) \eta_{\mu \nu}$, with $ \eta $ the conformal
cosmological time and $\eta_{\mu \nu}$ the components of the
Minkowskian metric, and $a(\eta)$ is the scale factor. During
slow-roll inflation, the scale factor can be approximated by $a(\eta)
\simeq -1/[H\eta(1-\epsilon_1)]$, with $H$ the Hubble factor, which
during inflation is approximately constant. The Hubble slow-roll
parameters, are defined as $\epsilon_1 \equiv 1- {\mH'}/{\mH^2},
\epsilon_2 \equiv {\epsilon_1'}/{\mH \epsilon_1}$, and both are very
small $\epsilon_1,\epsilon_2 \ll 1$. Here $\mH \equiv aH$ is the
conformal Hubble factor. Furthermore, in context of the slow-roll
approximation the slow-roll parameters are related to the inflaton
potential through,
\begin{subequations}\label{PSR}
\beq
\epsilon_1 \simeq \frac{M_P^2}{2} \left( \frac{\partial_\phi V}{V}\right)^2, 
\eeq
\beq
\epsilon_2 \simeq 2 M_P^2  \left[  \left( \frac{\partial_\phi V}{V}\right)^2 -  \left( \frac{\partial_{\phi \phi} V}{V}\right) \right].
\eeq
\end{subequations}
Within the slow-roll approximation, the equation of motion for the
background field is $3 \mH \phi_0' = -a^2 \partial_\phi V$.

The standard procedure is to split the scalar field and the metric
into background plus perturbations, i.e. $g_{ab} = g_{ab}^{(0)} +
\delta g_{ab}$ and $\phi (\vx,\eta) = \phi_0 (\eta) + \dphi
(\vx,\eta)$. At leading order in the scalar perturbations of the
background metric,{\footnote{In recent works
  {\cite{LKL15,bmodescsl1,bmodescsl2}}, we have focused on the tensor
  perturbations of the metric in the context of the semi-classical
  approach used in this paper. The results of those works indicate
  that the corresponding tensor modes are strongly
  suppressed. Therefore, in this paper, we only consider scalar
  perturbations of the metric.}} assuming no anisotropic stress, and
working in the longitudinal gauge, imply that the line element
associated to the scalar metric perturbations is
\begin{equation}
  ds^2 = a^2(\eta) \left[ - (1-2\Psi) d\eta^2 + (1-2 \Psi) \delta_{ij} dx^i dx^j \right].
\end{equation}

Before addressing the modified quantum dynamics and its impact in the
treatment of the infationary cosmology, we present the framework that
underlies our characterization of space-time metric and that of the
inflaton field.\footnote{For a further detailed presentation and
  motivation for our approach see ref.~\cite{PSS06,Shortcomings}.}
This framework is based on the semiclassical gravity approach, in
which the matter fields are treated quantum mechanically while the
gravity is treated in classical terms.\footnote{The point of view
  accepts that spacetime is quantum mechanical at the fundamental
  level, but considers that by the time that a metric characterization
  is meaningful, one is already well within the classical realm as
  far as the gravitational degrees of freedom are concerned. This view
  is quite natural once one considers, say the problem of time in
  canonical quantum gravity, and the regimes in which a notion of time
  might effectively emerge.} Note that this is a distinct view from
the standard approach in which the perturbations of both the metric
and the matter fields are treated in quantum mechanical terms.  The
framework we employ is thus based on Einstein's semiclassical
equations,
\begin{equation}
  G_{ab} = 8 \pi G \bra \hat T_{ab} \ket.
\end{equation}

We must mention the fact that direct calculations indicate the quantum
uncertainties in $\hat{T}_{ab}$ in this situation are in principle
very large \cite{Kuo1993} (in fact strictly speaking and when
considered at a given spacetime point these would be infinite), and
that might be taken as casting doubts about the validity not only of
semiclassical gravity but also of any kind of perturbative approach
underlying all treatments of cosmological perturbations. Nevertheless,
the fact that one obtains reasonable results, indicates the problem is
not insurmountable, and that some cutoff mechanism must be at
play.\footnote{For instance in \cite{bmodescsl1,bmodescsl2} it was
  argued that when considering the spectrum at the end of inflation,
  it was natural to take a cutoff scale to be given by the last scale
  that exits the horizon during inflation. On the other hand, when one
  is interested in comparing the theoretical predictions with the data
  from the CMB, one must take into account plasma damping effects and
  thus introduce a cutoff scale corresponding to the scale of
  diffusion or Silk damping} It should be noted however that in the
assessment of these issues one should be careful to compare the
uncertainties in the energy momentum tensor with the expectation value
of a full energy momentum tensor (a quantity that is extremely large
during inflation) and not just that of the space dependent
perturbations.  The problem however clearly deserves further study.

In our approach the initial state of the quantum field is taken to be
the same as the standard one, namely the Bunch-Davies
vacuum. Nevertheless, the self-induced collapse will spontaneously
change this initial state into a final one that does not need to share
the symmetries of the Bunch-Davies vacuum. Henceforth, after the
collapse $\bra \hat T_{ab} \ket$ will not have the symmetries of the
initial state, and this will led through Einstein semiclassical
equation, to a geometry, that. generically, will no longer be
homogeneous and isotropic. In particular, focusing on the metric
scalar perturbation $ \Psi$ Einstein's semiclassical equations in
Fourier space, at first order in the perturbation theory, led to:
\beq\label{master0} \Psi_{\nk} (\eta) = \sqrt{\frac{\epsilon_1}{2}}
\frac{H}{M_P k^2} a \bra \hat \dphi'_{\nk} (\eta) \ket, \eeq where
$M_P^2\equiv1/8 \pi G$ the reduced Planck mass.  Considering an
homogeneous and isotropic vacuum state for the field would lead to
$\bra 0 | \hat \dphi'_{\nk} |0\ket =0 $. It follows from
eq.~(\ref{master0}) that in the vacuum state $\Psi =0$ and
consequently the spacetime is perfectly homogeneous and isotropic. It
is only after after the self-induced collapse of the wave function,
associated to each mode of the inflaton, that $\bra \hat \dphi'_{\nk}
\ket \neq 0$, giving rise to the emergence of the primordial curvature
perturbations.

Note that,  in the standard treatment, there is no analogous expression to Eq. \eqref{master0}. In fact, the usual treatment is based on the quantization of both $\Psi$ and $\dphi$, which then are linearly combined in what is called the Mukhanov--Sasaki variable $v$ \cite{brandenberger1993}. Essentially,  the   treatment starts   with  the action at second order in perturbations, in a quasi--de Sitter spacetime background, expressed  in terms of the variable $v$. Such an action is then expanded  in Fourier modes, which takes the form  (for each mode)  of a harmonic oscillator with a ``time--dependent mass''. This  is   followed  by the canonical quantization of $v$. In the comoving gauge the variable $v$ and the curvature   perturbation $\mR$ (i.e the intrinsic spatial curvature on hypersurfaces on constant conformal time for a flat universe) are related by $\mR = v/z$, with $z \equiv a \sqrt{2 \epsilon_1} M_P$. Hence, a quantization of $v$ implies a quantization of $\mR$. Additionally,  the traditional approach assumes  that  when the proper wavelength of the mode becomes larger than the Hubble radius, a   certain  quantum to classical transition  takes place,  which  might be expressed  as   $\hat \mR_{\nk} \to \mR_{\nk}$  (the justification   for  assuming such   transition usually relies on arguments based on decoherence, the evolution of the vacuum state into a squeezed state, etc.). That is the quantum operator $\hat \mR_{\nk}$ is now taken as a classical stochastic field $\mR_{\nk} = A(k) e^{i \theta_{\nk}} $. The quantity $\theta_{\nk}$ is  a random phase, and $A(k)$ is a Gaussian random variable with zero mean and variance directly identified with the quantum uncertainty $\bra 0 | \hat{\mR}_{\nk}^2    | 0 \ket $. That is, the  focus of  the standard approach is the two--point quantum correlation function  $\bra 0 | \hat{\mR}_{\nk} \hat{\mR}_{\nk'}^\dagger | 0 \ket \equiv P(k) \delta(\nk-\nk')$ from where the power spectrum $P(k)$ is extracted.

In the next subsection, we will  focus on the  strictly semiclassical  approach\footnote{ Such approach is  sometimes  considered as  not viable,  but  as the    discussion illustrated    in \cite{unviable-semicalsssical1,unviable-semicalsssical2,unviable-semicalsssical3,unviable-semicalsssical4,unviable-semicalsssical5,unviable-semicalsssical6,unviable-semicalsssical7} the  arguments  are not  decisive. In particular, an  implementation involving a self--induced collapse  seems to be completely viable  as  far  as  mathematical consistency  and  phenomenology are concerned \cite{semicalsssical-and-collpase1,semicalsssical-and-collpase2}, and at least when regarding these  as effective theories.  } based  on quantum treatment of the matter fields  and
describe the modified dynamics corresponding to the self-induced
collapse. There are two main approaches that will be considered in
this paper: i) the one in which no particular collapse mechanism is
considered, and the collapse process is simply characterized in a
phenomenologically inspired scheme \cite{US08,LSS12,LLP15}; ii) a
second approach where a modification of the Schr\"odinger equation of
the CSL type leads naturally to the eventual collapse of the wave
function \cite{CPS13}. As mentioned in the introduction, we refer to
the first as the \emph{collapse schemes} approach, and the second one
as the continuous spontaneous localization (CSL) inflationary
approach.

\subsection{ Quantum treatment   and  Collapse schemes}\label{esquemas-colapso}

  The staring  point of the  treatment is
  the quantum theory of $\dphi (\vx,\eta)$ in a curved background
  described by a quasi-de Sitter space-time
  \cite{US08,LSS12,LLP15}. Moreover, it is convenient to work with the
  rescaled field variable $y=a\dphi$.  Both the field $y$ and the
  canonical conjugated momentum $\pi \equiv \partial \delta
  \mathcal{L}^{(2)}/\partial y' = y'-(a'/a)y=a\dphi'$ are promoted to
  quantum operators so that they satisfy the following equal time
  commutator relations: $[\hat{y}(\vx,\eta), \hat{\pi}(\vx',\eta)] =
  i\delta (\vx-\vx')$ and $[\hat{y}(\vx,\eta), \hat{y}(\vx',\eta)] =
  [\hat{\pi}(\vx,\eta), \hat{\pi}(\vx',\eta)] = 0$. Next, we can expand
  the field operator in Fourier modes, \beq \hat{y}(\eta,\vx) =
  \frac{1}{L^3} \sum_{\nk} \hat{y}_{\nk} (\eta) e^{i \nk \cdot \vx},
  \eeq with an analogous expression for $\hat{\pi}(\eta,\vx)$. Note
  that the sum is over the wave vectors $\vec k$ satisfying $k_i
  L=2\pi n_i$ for $i=1,2,3$ with $n_i$ integer and $\hat y_{\nk}
  (\eta) \equiv y_k(\eta) \ann_{\nk} + y_k^*(\eta) \cre_{-\nk}$ and
  $\hat \pi_{\nk} (\eta) \equiv g_k(\eta) \ann_{\nk} + g_{k}^*(\eta)
  \cre_{-\nk}$, with $g_k(\eta) = y_k'(\eta) - \mH y_k (\eta)$.  The
  equation of motion for the modes reads \beq\label{ykmov2}
  y''_k(\eta) + \left(k^2 - \frac{2+3\epsilon}{\eta^2} \right)
  y_k(\eta)=0, \eeq with $\epsilon \equiv -\epsilon_1 + \epsilon_2/2$.
  The selection of $y_k(\eta)$ reflects the choice of a vacuum state
  for the field. We proceed as in standard inflationary models and
  choose the so-called Bunch-Davies vacuum: 
  \beq\label{nucolapso} y_k (\eta) = \left( \frac{-\pi \eta}{4}
  \right)^{1/2} e^{i[\nu + 1/2] (\pi/2)} H^{(1)}_\nu (-k\eta), \eeq
  where $ \nu \equiv 3/2 + \epsilon$ and $H^{(1)}_\nu (-k\eta)$ is the
  Hankel function of the first kind of order $\nu$.
%

  The collapse hypothesis assumes that at a certain time $\tc$ the
  part of the state characterizing the mode $k$ randomly ``jumps'' to
  a new state, which is no longer homogeneous and isotropic. The
  collapse is considered to operate similar to an imprecise
  ``measurement,'' even though there is no external observer or
  detector involved. Therefore, it is reasonable to consider Hermitian
  operators, which are susceptible of a direct measurement in ordinary
  quantum mechanics.  Hence, we separate $\hat y_{\nk} (\eta)$ and
  $\hat \pi_{\nk} (\eta)$ into their real and imaginary parts $\hat
  y_{\nk} (\eta)=\hat y_{\nk}{}^R (\eta) +i \hat y_{\nk}{}^I (\eta)$
  and $\hat \pi_{\nk} (\eta) =\hat \pi_{\nk}{}^R (\eta) +i \hat
  \pi_{\nk}{}^I (\eta)$, such that the operators $\hat y_{\nk}^{R, I}
  (\eta)$ and $\hat \pi_{\nk}^{R, I} (\eta)$ are Hermitian
  operators. Thus,
\begin{subequations}\label{operadoresRI}
\beq
 \hat{y}_{\nk}^{R,I} (\eta) = \sqrt{2} \textrm{Re}[y_k(\eta) \hat{a}_{\nk}^{R,I}],
\eeq
\beq
\hat{\pi}_{\nk}^{R,I} (\eta) = \sqrt{2} 
\textrm{Re}[g_k(\eta) \hat{a}_{\nk}^{R,I}],
\eeq
\end{subequations} 
where $\hat{a}_{\nk}^R \equiv (\hat{a}_{\nk} + \hat{a}_{-\nk})/\sqrt{2}$, 
$\hat{a}_{\nk}^I \equiv -i (\hat{a}_{\nk} - \hat{a}_{-\nk})/\sqrt{2}$. 

The commutation relations for the $\hat{a}_{\nk}^{R,I}$ are
non-standard: \beq\label{creanRI}
[\hat{a}_{\nk}^{R,I},\hat{a}_{\nk'}^{R,I \dag}] = L^3
(\delta_{\nk,\nk'} \pm \delta_{\nk,-\nk'}), \eeq the $+$ and the $-$
sign corresponds to the commutator with the $R$ and $I$ labels
respectively; all other commutators vanish.  It is also important to
emphasize that the vacuum state defined by $ \ann_{\nk}{}^{R,I} |0\ket
=0$ is fully translational and rotationally invariant (see the formal
proof in Appendix A of ref.~\cite{LLS13}).

%

Next, we need to specify the dynamics of the expectation values $\bra
\hat{y}^{R, I}_{\nk} (\eta) \ket$ and $\bra \hat{\pi}^{R, I}_{\nk}
(\eta) \ket$, evaluated in the post-collapse state, which will depend
on the expectation values evaluated at the time of collapse of each
mode of the field $\tc$. In the \emph{collapse schemes} approach, we
do not consider a specific collapse mechanism, instead we characterize
the post-collapse state by the expectation value and the quantum
uncertainty of the fields at the time $\tc$. In the present work, we
will consider only two possibilities for such relations. Namely, the
\emph{Newtonian} and \emph{Wigner} collapses schemes analyzed in
ref.~\cite{LLP15}.  We do not consider the \emph{independent} scheme
studied in the same work since it has been shown \cite{LSS12,LLP15}
that when considering the collapse time $\eta_c^k= \frac{A}{k}+B$, the
CMB angular spectrum, associated to that scheme, is indistinguishable
from the prediction of the standard inflationary model.

\subsubsection{Newtonian collapse scheme}

This scheme is motivated by the fact that only the expectation value of the momentum operator $\hat \pi_{\nk} \equiv a \hat \dphi'_{\nk}$ appears as a source for the curvature perturbation $\Psi_{\nk}$ in Eq. \eqref{master0}. Also, this view seems to be close in spirit to the ideas of Penrose \cite{penrose1996} regarding quantum uncertainties that the gravitational potential would be inheriting from the matter fields' quantum uncertainties. Therefore, in this scheme the collapse affects only the expectation value of the
conjugated momentum variable, i.e.,

\beq\label{esquemanewt}
 \bra \hat{y}^{R,I}_{\nk}(\eta^c_{\nk})\ket  = 0, \qquad
  \bra \hat{\pi}^{R,I}_{\nk}(\tc) \ket = x_{\nk,2}^{R,I}
  \sqrt{\left(\Delta \hat{\pi}^{R,I}_{\nk} (\tc) \right)^2_0}.
\end{equation}
where, $x_{\nk,2}^{(R,I)}$ represents a random Gaussian 
variable normalized and centered at zero. The quantum uncertainties of the vacuum state at the time of collapse are \cite{LLP15}:
\begin{equation}
  \left( \Delta \hat{y}^{R,I}_{\nk} (\tc) \right)^2_0 = \frac{L^3 \pi |z_k|}{16 k} \left[ J_\nu^2 (|z_k|) + Y_\nu^2 (|z_k|) 
  \right],
 \end{equation} 
\begin{eqnarray}
  \left(\Delta \hat{\pi}^{R,I}_{\nk} (\tc) \right)^2_0 &=& \frac{L^3 \pi k }{16} \bigg[ \bigg( \frac{-\alpha 
    J_\nu (|z_k|)}{\sqrt{|z_k|}} \nn
  &+& \sqrt{z_k|} J_{\nu+1} (|z_k|) \bigg)^2 + \bigg( \frac{-\alpha Y_\nu (|z_k|)}{\sqrt{|z_k|}} \nn
  &+& \sqrt{|z_k|} Y_{\nu+1} (|z_k|) \bigg)^2 \bigg], 
\end{eqnarray} 
where $\alpha \equiv 1/2 + \nu$, $J_\nu $ and $Y_\nu$ are the Bessel
functions of the first and second kind respectively; $z_k \equiv k
\tc$ and $\tc$ is the time of collapse for each mode.

\subsubsection{Wigner collapse scheme}

Heissenberg's uncertainty principle indicates that quantum uncertainties of position and momentum operators are  not independent. In particular, momentum and position of a quantum system cannot  be determined independently and simultaneously. As we have mentioned, the self-induced collapse acts as a sort of  spontaneous  ``measurement" (of course without relying on observers or measurements devices) of some variable  involving both position and momentum. Therefore,  as suggested  by the uncertainty principle,  the collapse might  involve   correlations between position and momentum.  Extrapolating this idea to our situation of interest indicates  that the self-induced collapse could correlate the field $\hat y$ and its conjugated momentum $\hat \pi$. One possible way to characterize the correlation is to use Wigner's distribution function. In non-relativistic quantum mechanics, Wigner's function can be considered, under certain special circumstances, as a probability distribution function for a quantum system, i.e. it allow us to visualize the momentum-position correlations and quantum interferences in ``phase space". For the vacuum state of each mode of the inflaton, the corresponding Wigner's function is a bi-dimensional Gaussian. As a consequence, in this scheme  we will characterize the post-collapse expectation values as: 
\begin{subequations}\label{esquemawig}
 \beq
\bra \hat{y}^{R,I}_{\nk}(\eta^c_{\nk})\ket  = x_{\nk}^{R,I} \Lambda_k (\tc) 
\cos \Theta_k (\tc), 
\eeq
\beq
  \bra \hat{\pi}^{R,I}_{\nk}(\tc) \ket = x_{\nk}^{R,I} k \Lambda_k (\tc) 
\sin \Theta_k (\tc),
\eeq
\end{subequations}
where $x_{\nk}^{R,I}$ is a random variable, characterized by a
Gaussian probability distribution function centered at zero with
spread one. The parameter $\Lambda_k (\tc)$ represents the major
semi-axis of the ellipse in the $\hat{y}-\hat{\pi}$ plane where the
Wigner function has 1/2 of its maximum value. The other parameter
$\Theta_k (\tc)$ is the angle between $\Lambda_k (\tc)$ and the
$\hat{y}_{\nk}^{R,I}$ axis. For details involving the Wigner function
and the collapse scheme we refer the reader to ref.~\cite{US08}. The
explicit expressions for $\Lambda_k$ \cite{LLP15} and $\Theta_k $ are
\barr\label{lambdak} &\Lambda_k& = (2L)^{3/2} \sqrt{\frac{\pi
    |z_k|}{4k}} \left[ J_\nu^2 (|z_k|) +
  Y_\nu^2 (|z_k|) \right]^{1/2} \bigg[ S(|z_k|)  \nonumber \\
&-& \sqrt{S^2 (|z_k|) - \left(\frac{\pi |z_k|}{2}\right)^2 (J_\nu^2
  (|z_k|) + Y_\nu^2 (|z_k|) )^2} \bigg]^{-1/2},\nn \earr
\barr\label{2thetak} \tan 2\Theta_k &=& -\frac{\pi^2 |z_k|}{4} \left[
  J_\nu^2 (|z_k|) + Y_\nu^2 (|z_k|) \right] \bigg[ S(|z_k|) \nn &-&
\frac{\pi |z_k|}{8} \left(
  J_\nu^2 (|z_k|) + Y_\nu^2 (|z_k|) \right)^2 \bigg]^{-1} \nonumber \\
&\times& \bigg[ -2\nu \left( J_\nu^2 (|z_k|) + Y_\nu^2 (|z_k) \right)
\nn &+& |z_k| \bigg( J_\nu (|z_k|) J_{\nu+1} (|z_k|) \nn &+& Y_\nu
(|z_k|) Y_{\nu+1} (|z_k|) \bigg) \bigg], \earr where \barr\label{defS}
S(|z_k|) & \equiv& 1 + \frac{\pi^2}{16} \bigg\{ |z_k|^2 (J_\nu^2
(|z_k|) + Y_\nu^2 (|z_k|))^2 \nn &+& 4 \bigg[ J_\nu^2 (|z_k|) +
Y_\nu^2 (|z_k|) - |z_k| ( J_\nu (|z_k|)
\nonumber \\
&\times& J_{\nu+1} (|z_k|) + Y_\nu (|z_k|) Y_{\nu+1} (|z_k|) )
\bigg]^2 \bigg\}.  \earr

\subsection{CSL inflationary approach}\label{CSLapproach}

The implementation of the CSL model into the slow-roll inflationary
model, using the semiclassical gravity framework, has been analyzed
originally in ref.~\cite{CPS13}. Here, we provide the main features of
the CSL inflationary model generalized into a quasi-de Sitter
spacetime background.

The CSL model is based on a modification of the Schr\"odinger
equation. This alteration induces a collapse of the wave function
towards one of the possible eigenstates of an operator called the
collapse operator with certain rate. The objective
reduction process is due to the interaction of the system with a
background noise, which is a continuous-time
stochastic process of the Wiener kind (see \cite{bassi2003,bassi2013} for a throughly
review). We will be more precise in the following.

In order to apply the CSL model to the inflationary setting, we will
follow the approach first introduced in \cite{CPS13}. That work relies
on a version of the CSL model in which the nonlinear aspects of the
CSL model are shifted to the probability law. That is, the evolution
law is linear just as the Schr\"odinger equation, but then, the law of
probability for the realization of a specific random function, becomes
dependent of the state that results from such evolution. Specifically,
the theory can be characterized in terms of two equations: The first
is a modified Schr\"odinger equation, whose solution is
\begin{equation}\label{CSLQM}
|\psi,t\rangle={\cal T}e^{-\int_{0}^{t}dt'\big[i\hat H+\frac{1}{4\lambda}[w(t')-2\lambda_0 \hat A]^{2}\big]}|\psi,0\rangle.
\end{equation}
$\cal T$ is the time-ordering operator, $w(t)$ characterizes the stochastic process, i.e. is a random classical
function of time, of white noise type. The modification of Schr\"odinger's equation given by the CSL model induces the collapse of the wave function towards one of the possible eigenstates of $\hat A$, that is, the operator $\hat A$ is the \textit{collapse operator}. In laboratory situations, the collapse operator is usually chosen to be the position operator \cite{bassi2013}.  The parameter $\lambda_0$ is the universal CSL parameter that serves to set the strength of the collapse.  The value of $\lambda_0$ characterizes the rate at which the wave function increases its ``localizations'' in the eigen-basis
	of the collapse operator.

The probability for the $w(t)$ is given by the second equation, the Probability Rule
\begin{equation}
	PDw(t)\equiv\langle\psi,t|\psi,t\rangle\prod_{t_{i}=0}^{t}\frac{dw(t_{i})}{\sqrt{ 2\pi\lambda/dt}}.
\end{equation}
In Ref. \cite{CPS13} it is shown that with the appropriate selection
of the field collapse operators and using the corresponding CSL
evolution law one obtains collapse in the relevant operators
corresponding to the Fourier components of the field and the momentum
conjugate of the field. This bypasses any concerns regarding possible
mode mixing at the first order in perturbation theory (at higher order
there is mode mixing even in the traditional treatments).\footnote{We
  also acknowledge at this point that there is no complete version of
  the CSL theory that is applicable in all situations, ranging from
  laboratory ones to the ones involving cosmology and black hole
  space-times. Nevertheless, we adopt the point of view that proposing
  educated guesses, in combination with phenomenological models
  applicable to particular situations, allow us to progress in our
  program. We think this is analogous to the path that took physics
  down the road that ended with the standard model of particle
  physics, namely trial and error focusing first on rather specific
  situations, and then looking for ways to generalize, based on what
  was found to work in each case.  }

Given that the CSL model modifies the Schr\"odinger equation, it is
convenient to describe the quantum theory of the inflaton in the
Schr\"odinger picture, where the relevant objects are the wave
function and the Hamiltonian.

The Hamiltonian characterizing the inhomogeneous sector of the
inflaton is $H = (1/2) \int d^3 k (H_{\nk}^\textrm{R} +
H_{\nk}^\textrm{I})$ with
\begin{equation}\label{hamilt}
   H_{\nk}^\RI = \pi_{\nk}^\RI \pi_{\nk}^{*\RI} + k^2 y_{\nk}^\RI y_{\nk}^{*\RI} 
   - \frac{(1-\epsilon_1+\epsilon_2/2)}{\eta} \left(  \yk^\RI \pk^{*\RI} +  \yk^{*\RI} \pk^{\RI}     \right)
\end{equation}

where $\yk = a \dphi_{\nk}$ and $\pi_{\nk} \equiv y_{\nk} '- \mH
y_{\nk} $. The indexes R,I denote the real and imaginary parts of
$\yk$ and $\pk$. We remind the reader that $\epsilon_1$ and
$\epsilon_2$ are the Hubble slow roll parameters defined at the
beginning of this section. We now promote $\yk$ and $\pk$ to quantum
operators, by imposing canonical commutations relations $[\hat
y_{\nk}^\RI, \hat \pi_{\nk}^\RI] = i \delta (\nk-\nk'). $

 In  order  to  apply the CSL theory to the situation at hand  we need  to make an educated  guess  regarding  the  collapse operator $\hat A$ that should   drive the modified   dynamics  in  this case.  As  we  explained in the case of non-relativistic quantum mechanics the  operator  $\hat A$ is taken as a smeared position operator,  which  could  be associated  with a   sort of mass density  (specially  if the collapse parameter is proportional to the  particle's mass   as  suggested in a previous work\cite{bassi2013}).  One might interpret that view  as  indicating that  the collapse  is tied  with some aspect of  the quantum matter that ``gravitates" (i.e. that would  characterize the  interaction between  gravitation and matter degrees of freedom).  Thus,   extrapolating that idea to the situation at hand,   we  can guess  that  we should look at the  quantity appearing in the relevant     component of  Einstein's  semiclassical equation  as a   candidate for the collapse operator.  Considering now the form  of the relevant  component of  such equation  given  by Eq. \ref{master0}  this  line of thought takes  us to consider  the  momentum conjugate to the field  as a  rather natural candidate.  For reasons mentioned at the beginning of this subsection, one may
apply the CSL reduction mechanism on each mode of the field
independently. That is,  we  assume  that the momentum operator $\hat{\pi}_{\nk}^{\textrm{R,I}}$  in each mode  acts  as the collapse
operator  for that mode.  However, as in  any such situation, the suitability  of  an  educated  guess  must  be decided  by the  long  term  empirical  success or failure of  the emerging  predictions.
Therefore, the evolution of the state vector  characterizing the  quantum field  as given by the CSL
theory  is:
\begin{equation}\label{CSLevolution}
  |\Phi_{\nk}^{\textrm{R,I}}, \eta \ket = \hat T \exp \bigg\{ - \int_{\tau}^{\eta} 
d\eta'   \bigg[ i \hat{H}_{\nk}^{\textrm{R,I}} 
+ \frac{1}{4 \lambda_k} (\mathcal{W}(\nk,\eta') - 2 \lambda_k
\hat{\pi}_{\nk}^{\textrm{R,I}})^2 \bigg] \bigg\} |\Phi_{\nk}^{\textrm{R,I}}, 
  \tau \ket,
\end{equation} 
$\hat T$ is the time-ordering operator, and $\tau$ denotes the
conformal time at the beginning of inflation. In addition, we have generalized the white noise $w(t)$ appearing in Eq. \eqref{CSLQM} into a stochastic field $\mathcal{W}$ which depends on $\nk$ and the conformal time. That is, since we are applying the CSL collapse dynamics to each mode of the field, it is natural to introduce a stochastic function for each independent degree of freedom. Henceforth, the stochastic field $\mathcal{W}(\nk,\eta)$ might be regarded as a Fourier transform on a stochastic spacetime field $\mathcal{W}(\vec x,\eta)$.

Given that we take  the momentum operator $\hat{\pi}_{\nk}^{\textrm{R,I}}$ to  act as the collapse operator, it is convenient to
 work with the wave function in the momentum representation. We denote
 by $\Phi[\pi]$ the wave function characterizing the quantum state of
 the field. In Fourier space, the wave function can be factorized into
 mode components $\Phi[\pi] = \Pi_{\nk} \Phi_{\nk}^\textrm{R}
 [\pi_{\nk}^\textrm{R}] \times \Phi_{\nk}^\textrm{I}
 [\pi_{\nk}^\textrm{I}] $. 
 
It is known that the ground state of the Hamiltonian \eqref{hamilt} characterized by a wave functional in the momentum representation $ \Phi^\RI_0 (\pk^\RI)$ is a Gaussian. Additionally, the Hamiltonian \eqref{hamilt} and the CSL evolution equation \eqref{CSLevolution} are quadratic in both $\hat{\pi}_{\nk}^{\textrm{R,I}}$ and $\hat{y}_{\nk}^{\textrm{R,I}}$; consequently, the form of the wave function at any time in the momentum basis is  \cite{CPS13}: 
%
%
%
%
 \begin{equation}\label{wf}
 \Phi^\RI (\eta,\pk^\RI) =
 \exp[-A_k(\eta) (\pk^\RI )^2 + B_k^\RI(\eta) \pk^\RI + C_k^\RI (\eta) ].
 \end{equation} 
 
The evolution equation \eqref{CSLevolution} when applied to the  wave
	functional \eqref{wf}, results in a set of dynamical equations for
the objects $A_k(\eta)$, $B_k^\RI(\eta)$, and $C_k^\RI (\eta)$. The
initial conditions are set by the initial state of the field, i.e. the
Bunch-Davies vacuum. That is, the initial conditions are $A_k(\tau) =
1/2k$, $B_k^\RI(\tau)=0$, and $C_k^\RI (\tau)=0$. As a matter of fact,
we are only interested in the equation of motion for $A_k(\eta)$ since
this quantity is directly related to the primordial spectrum.  The
analysis of \cite{CPS13} in fact indicates that,
 \begin{equation}
  A_k' = \frac{i}{2} + \lambda_k - 2A_k \frac{(1-\epsilon_1+\epsilon_2/2)}{\eta} -2ik^2 A_k^2.
 \end{equation} 
The solution of the latter equation is
\begin{equation}\label{Ak}
  A_k (\eta) = \frac{q}{2ik^2} \left[\frac{ J_{\mu+1} (-q\eta) + e^{-i \pi \mu} J_{-\mu-1} (-q\eta)   }{J_\mu (-q\eta) - e^{-i \pi \mu} J_{-\mu} (-q\eta)  } \right],
\end{equation} 
with $q^2 \equiv k^2 (1-2i\lambda_k)$ and $ \mu \equiv 1/2 -\epsilon_1 + \epsilon_2/2$. 


\section{Primordial Power Spectrum for the collapse models}
\label{primordialpower}

In this section, we briefly review the procedure to obtain the
primordial scalar power spectrum for the collapse approaches described
in the previous section. Afterwards, in the following sections, we
will compare the predictions for the primordial power spectra
resulting from the collapse models, with recent CMB data.


We begin by characterizing the CMB radiation in terms of the
temperature anisotropies $\Theta (\hat {n}) \equiv \delta T/T_0$ of
the CMB, with $T_0$ the mean temperature and $\delta T \equiv T
\hat{n} - T_0$ where $ T \hat{n}$ is the temperature of the CMB
radiation in the direction $\hat{n}$ in the sky. The coefficients
$a_{lm}$ of the spherical harmonic expansion of $\Theta (\hat {n})$
are

\beq\label{alm0} a_{lm} = \int \Theta (\hat n) Y_{lm}^\star
(\theta,\varphi) d\Omega, \eeq with $\hat n = (\sin \theta \sin
\varphi, \sin \theta \cos \varphi, \cos \theta)$ and $\theta,\varphi$
the coordinates on the celestial two-sphere.  The Fourier
decomposition for the temperature anisotropies can be written as
follows: $\Theta (\hat n) = \sum_{\nk} ({\Theta (\nk)}/{L^3}) e^{i \nk
  \cdot R_D \hat n}$ with $R_D$ being the radius of the last
scattering surface. We recall that $\Theta (\hat n)$ is directly
related to the primordial curvature perturbation. In the comoving
gauge, which is the one considered in the numerical code we are going
to use in the next section, the curvature perturbation is given by the
field $\mR$.

In Fourier space, the temperature anisotropies and the initial
curvature perturbation are related as $\Theta (\nk) = T(k) \mR_{\nk}$,
where $T(k)$ is the transfer function which contains the physics
between the beginning of the radiation-dominated era and the
present. Consequently, the coefficients $a_{lm}$, in terms of the
modes $\mR_{\nk}$ can be expressed: \beq\label{alm2} a_{lm} = \frac{4
  \pi i^l}{ L^3} \sum_{\nk} j_l (kR_D) Y_{lm}^\star(\hat k) T (k)
\mR_{\nk}, \eeq with $j_l (kR_D)$ being the spherical Bessel function
of order $l$.

On the other hand, in the \emph{collapse schemes} and the CSL
inflationary approaches, the theoretical predictions were obtained
choosing the longitudinal gauge. In that gauge, the curvature
perturbation is characterized by the Newtonian potential $\Psi$.  The
relation between $\Psi$ and $\mR$ is $\mR = \Psi + (2/3)(\mH^{-1}
\Psi' + \Psi )/(1+\omega)$, with $\omega \equiv P/\rho $
\cite{brandenberger1993}\cite{Mukhanov:1990me}. During the inflationary epoch $\omega + 1
\simeq (2/3) \epsilon_1$. In fact, for the modes of observational
interest $\mR_{\nk} \simeq \Psi_{\nk} / \epsilon_1$, with $\Psi_{\nk}$
given in eq.~(\ref{master0}). Hence, eq.~(\ref{alm2}) can be recasted
as,
\begin{equation}\label{alm3}
  a_{lm} = \frac{4 \pi i^l}{ L^3} \sum_{\nk} j_l (kR_D) Y_{lm}^\star(\hat k) T (k) \frac{\Psi_{\nk}}{\epsilon_1}.
\end{equation} 
Furthermore, using eq.~(\ref{master0}) and the definition of the
conjugated momentum of the field $\hat y$, the expression for the
coefficients $a_{lm}$ can be expressed in the final form
\begin{equation}\label{alm4}
  a_{lm} = \frac{4 \pi i^l}{ L^3} \frac{H}{\sqrt{2\epsilon_1} M_P }  \sum_{\nk} j_l (kR_D) Y_{lm}^\star(\hat k) T (k) 
  \frac{\bra \hat \pi_{\nk}  \ket}{k^2}.
\end{equation}

In the \emph{collapse schemes} and CSL inflationary approaches, the
expectation value $\bra \hat \pi_{\nk} \ket$ is a random variable: in
the first such random variable is characterized by $x_{\nk}$, while in
the CSL inflationary approach, each realization of $\bra \hat
\pi_{\nk} \ket$ corresponds to a particular post-collapse state, the
stochasticity of said state is generated from the noise function
$\mathcal{W}$. As a consequence, the coefficients $a_{lm}$, given in
eq.~(\ref{alm4}), are, in effect, a sum of random complex numbers,
just like in an effective two-dimensional random walk. Nevertheless,
one cannot give a perfect estimate for the direction of the final
displacement resulting from the random walk, instead, one might
provide an estimate for the length of the displacement. Thus, we can
obtain an estimate for the most likely value of $|a_{lm}|^2$, and
interpret it as the theoretical prediction for the observed
value. Moreover, such estimate can be obtained as follows: given that
the collapse is characterized by a random process, we can consider a
set of possible realizations of such process leading in each case to a
specific universe. That is, we consider an imaginary ensemble of
universes, each member of the ensemble is characterized by the set of
random variables $\bra \hat \pi_{\nk} \ket$ for all $ {\nk} $. If we
assume no correlation between different modes, and approximate the
distribution of $\bra \hat \pi_{\nk} \ket$ by a Gaussian, then we can
identify the most likely value $|a_{lm}|^2_{\text{ML}}$ with the mean
value $\overline{|a_{lm}|^2}$ of all possible realizations, i.e.,
$|a_{lm}|^2_{\text{ML}}= \overline{|a_{lm}|^2}$.

The quantity that is used in the statistical analysis to compare with
observational data is the angular power spectrum: $C_l = (2l+1)^{-1}
\sum_m |a_{lm}|^2$. The previous discussion lead us to identify the
observed value of $|a_{lm}|^2$ with the ensemble average $
\overline{|a_{lm}|^2}$. Hence, after passing to the continuum, the
theoretical angular power spectrum is
\begin{equation}\label{clfinal}
  C_l = 4 \pi \int_0^\infty \frac{dk}{k} j_l(k R_D)^2 T(k)^2 P(k),
\end{equation} 
where $P(k)$ is a function of $k$ that can be interpreted as an
\emph{effective power spectrum} (dimensionless), which is given by
\begin{equation}\label{PS0}
  P(k) = \frac{H^2}{k M_P^2 \epsilon_1} \overline{ \bra \hat \pi_{\nk} \ket \bra \hat \pi_{\nk} \ket^* }  
\end{equation} 

In the latter equation it is clear that the effective power spectrum
is not the same as the one in the standard approach. Indeed the latter
is identified with the quantum two-point correlation function $\bra 0
| \hat \mR(x) \hat \mR(y) | 0 \ket$, while the former, is obtained
from the ensemble average of two-product expectation values $\bra \hat
\pi_{\nk} \ket$ in the post-collapse state. The explicit form of the
effective power spectrum depends on whether the \emph{collapse
  schemes} or the CSL inflationary approach is being used.\footnote{ A
  more technical presentation on how to obtain the effective power
  spectrum and its conceptual meaning is given in Appendix D of
  ref.~\cite{LLP15}.}

\subsection{Effective power spectrum in the \emph{collapse schemes} approach}

In the \emph{collapse schemes} approach, the evolution of the
expectation value $\bra \hat \pi_{\nk} (\eta) \ket$, is calculated in
terms of the expectations values $\bra \hat \pi_{\nk} (\tc) \ket$ and
$\bra \hat y_{\nk} (\tc) \ket$ evaluated at the time of collapse of
the mode $\vec k $ : $\tc$ (all the expectations values are taken in
the post-collapse state). In particular, one obtains an expression of
the form
\begin{equation}\label{expecpieta}
  \bra \hat \pi_{\nk} (\eta) \ket = F(k\eta,z_k) \bra \hat y_{\nk} (\tc) \ket + G(k\eta,z_k)  \bra \hat \pi_{\nk} (\tc) \ket, 
\end{equation} 
where we recall that $z_k \equiv k \tc$ while the explicit expressions
for the functions $F$ and $G$ are given in ref.~\cite{LLP15}. The
expectation values $\bra \hat \pi_{\nk} (\tc) \ket$ and $\bra \hat
y_{\nk} (\tc) \ket$ are characterized for each collapse scheme. In the
\emph{Newtonian} and \emph{Wigner} schemes, the proposed
characterization is shown in eqs.~\eqref{esquemanewt} and
\eqref{esquemawig} respectively.

Given that the transfer functions $T(k)$ involved in the final
expression for the angular spectrum, eq.~\eqref{clfinal}, encode the
post-inflationary evolution of the primordial perturbations, we
evaluate $ \bra \hat \pi_{\nk} (\eta) \ket $ at a time near the end of
the inflationary regime, i.e. when $-k \eta \ll 1$. On the other hand,
within our assumptions the collapse can take place at any time during
inflation. In particular, it can occur when the proper wavelength of
the mode is bigger or smaller than the Hubble radius.  In this paper,
we focus on the case where the proper wavelength associated to the
mode is smaller than the Hubble radius, at the time of collapse; in
other words $k \gg a(\tc) H$, which is equivalent to $-k \tc \gg 1$.

Therefore, using eqs.~\eqref{PS0} and \eqref{expecpieta}, the
equivalent power spectrum that results from the \emph{collapse
  schemes} approach is \cite{LLP15}
\begin{equation}\label{PSesquemasfinal}
  P(k) = A_s Q(z_k) k^{n_s-1}.
\end{equation} 
The predicted amplitude of the power spectrum is similar to the one
given in the standard inflationary picture, $A_s \propto H^2/M_P^2
\epsilon_1$. However, the collapse hypothesis modifies the prediction
for the spectral index
\begin{equation}
 n_s - 1 = 2 \epsilon_1 - \epsilon_2
\end{equation} 

We recall that in the standard inflationary scenario: $n_s - 1 = -2
\epsilon_1 - \epsilon_2$ .  In addition, there is a new function of
the time of collapse $Q(z_k)$ which is different for each collapse
scheme; in the \emph{Newtonian} scheme,
\begin{eqnarray}\label{pnewtdentro}
& &  Q^{\textrm{Newt.}}(z_k) = \left[  1 + 
\frac{1}{|z_k|^2} \left( -2\nu + \frac{\Gamma(\nu + 5/2)}{2 \Gamma(\nu + 1/2)}  
\right)^2       \right] \nn
&\times& \bigg[ \cos \beta(\nu,|z_k|) - \frac{\sin \beta(\nu, |z_k|) }{2|z_k|} 
\frac{\Gamma(\nu+3/2)}{\Gamma(\nu - 1/2)}     \bigg]^2 \nn
\end{eqnarray} 
while in the \emph{Wigner} scheme, 
\begin{eqnarray}\label{pwignerdentro}
 & & Q^{\textrm{Wig.}}(z_k) = 
 \bigg\{  \bigg[ \frac{2\nu}{\zk^{3/2}}  \left( \cos \beta(\nu,|z_k|) - \frac{\sin \beta(\nu, 
|z_k|) }{2|z_k|} \frac{\Gamma(\nu+3/2)}{\Gamma(\nu - 1/2)}     \right)  \nn
&-& \left( \sin \beta(\nu,|z_k|) + \frac{\cos \beta(\nu, |z_k|) }{2|z_k|} 
\frac{\Gamma(\nu+5/2)}{\Gamma(\nu + 1/2)}     \right)            \bigg] \cos 
\Theta_k   \nonumber \\
&+&  \left[ \cos \beta(\nu,|z_k|) - \frac{\sin \beta(\nu, |z_k|) }{2|z_k|} 
\frac{\Gamma(\nu+3/2)}{\Gamma(\nu - 1/2)}     \right] \sin \Theta_k  \bigg\}^2, \nn
\end{eqnarray} 
where  $\nu=2 - {n_s}/{2}$, $\beta(\nu,|z_k|) \equiv |z_k| - (\pi/2) (\nu+1/2)$ and $\tan 2\Theta_k  \simeq -4/3\zk$.

It follows from eq.~(\ref{PSesquemasfinal}) that if we consider $z_k$
independent of $k$, then we recover the standard shape of the
spectrum, that is $P(k)^{\textrm{std.}} \propto k^{n_s-1}$
. Furthermore, in previous works \cite{LSS12,LLP15}, small departures
from this expression of the form $z_k = \mathcal{A} + \mathcal{B} k$,
were considered. For this choice of $z_k$ the collapse time of each
mode reads:
\beq
\eta_c^k = \frac{\mathcal{A}}{k}+\mathcal{B}, 
\label{ctime}
\eeq
where \A is dimensionless and \B has units of {\rm Mpc}. The
comparison between the primordial power spectrum, which resulted from
the \emph{Newtonian}/\emph{Wigner} schemes, and the standard spectrum
from the traditional inflationary model has been shown and discussed
thoroughly in ref.~\cite{LLP15} for different values of \A and \B. A
statistical analysis contrasting the effect of this kind of dependence
of collapse time on the mode's wave number on the CMB spectrum with
WMAP9 data has been performed in ref.~\cite{LSS12}.  On the other
hand, in a recent work, some of us have also studied a different
possibility for the dependence of collapse time on the mode's wave
number which affects predominantly the low $\ell$ part of the CMB
spectrum \cite{BLA16}. Results from a Bayesian model comparison
analysis indicate that the data show no preference between the Wigner
collapse model and the standard $\Lambda$CDM model. Therefore, it is
interesting to analyze the effect of the dependence in
eq.~\eqref{ctime} with recent CMB data and perform a Bayesian model
comparison analysis.  Here, we mention that the inflationary expansion
period 
corresponds to negative conformal time, so we choose to work with
negative values for \A and \B.

\begin{figure}
\begin{center}
\includegraphics[scale=0.7]{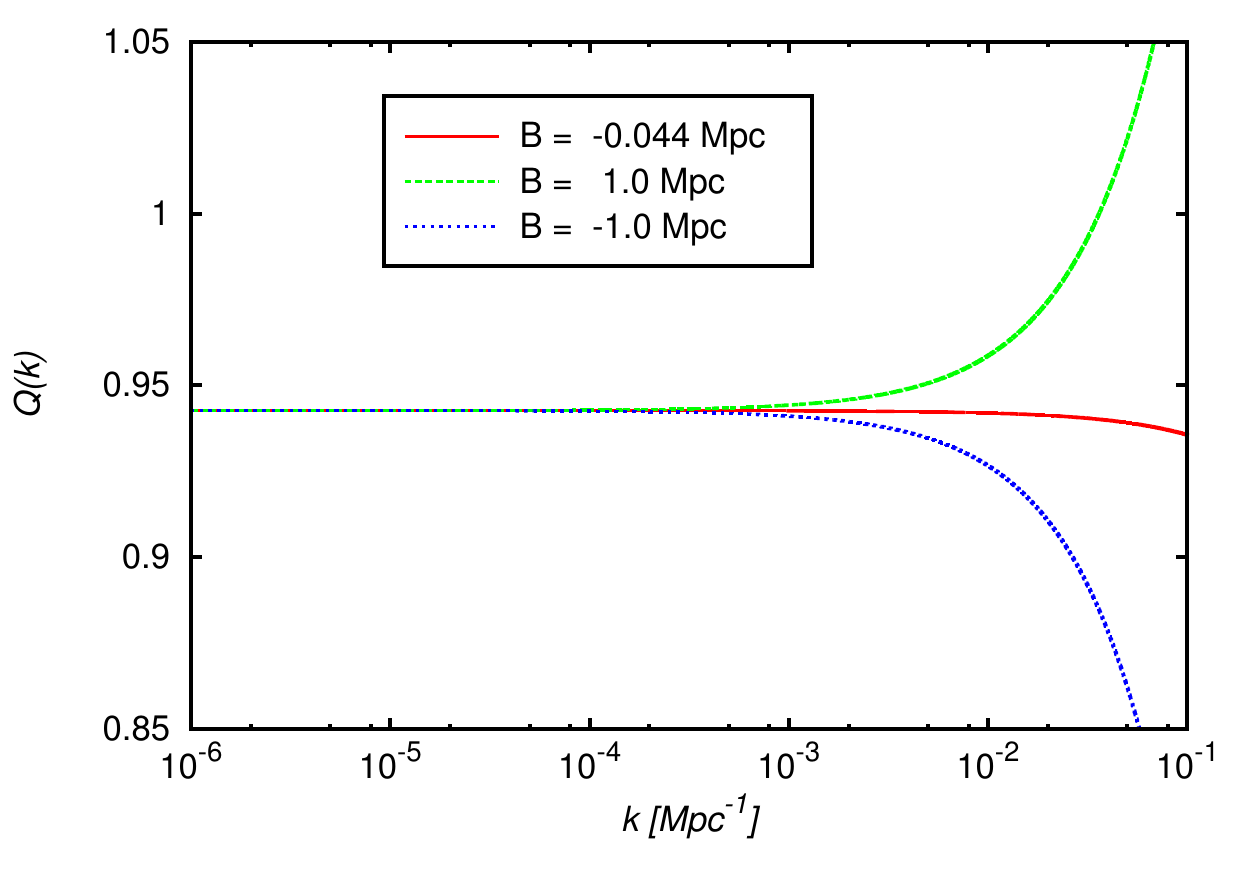}
\end{center}
\caption{The function $Q(k)$ associated to the power spectrum in the
  \emph{Wigner} collapse scheme. We have set the value \A$= -750$. We
  have considered three different values of \B (displayed in the
  figure).}
\label{figCkWigner}
\end{figure}

Figure \ref{figCkWigner}, shows $Q(k)$, the modification of the power spectrum in the  \emph{Wigner} scheme for three different values of \B and fixed \A. Recall that the $k$ dependence on $Q$ is inherited through the variable $z_k \equiv k \tc$, see Eqs. \eqref{pwignerdentro} and \eqref{ctime}. Also, it should be noted that $Q(k) =$ constant means no
change in the standard shape of the power spectrum. The effect of
considering the \emph{Wigner} collapse scheme  on the primordial power spectrum   induces an important modification in both
the amplitude and shape of the mentioned spectrum.  Besides, the intensity of the change depends on the value of \B. It follows from Figure  \ref{figCkWigner} that the influence of the collapse scheme is most significant for high values of $k$, which will result in a change in the small angular scales of the CMB temperature  and polarization spectrum (see Fig. \ref{clswigner}). The same analysis can be done for the \emph{Newton} collapse scheme, resulting in similar conclusions (see Ref. \cite{LLP15}).

\subsection{Effective power spectrum in the CSL inflationary approach}

As indicated by eq.~\eqref{PS0}, one needs to compute the average $
\overline{ \bra \hat \pi_{\nk} \ket \bra \hat \pi_{\nk} \ket^* } $ in
order to provide a prediction for the effective power spectrum. That
average is
\begin{eqnarray}
  \overline{ \bra \hat \pi_{\nk} \ket \bra \hat \pi_{\nk} \ket^* } &=& \overline{\bra \hat \pi_{\nk}^\textrm{R} + i \hat \pi_{\nk}^\textrm{I}\ket \bra \hat \pi_{\nk}^\textrm{R} - i \hat \pi_{\nk}^\textrm{I} \ket} \nn
  &=& \overline{\bra \hat \pi_{\nk}^\textrm{R} \ket^2} + \overline{\bra \hat \pi_{\nk}^\textrm{I} \ket^2}.
\end{eqnarray} 
Furthermore, $ \overline{ \bra \hat \pi_{\nk}^\textrm{R} \ket^2 }= \overline{\bra \hat \pi_{\nk}^\textrm{I} \ket^2}$. In the CSL inflationary approach, it can be shown \cite{CPS13} that,
\begin{equation}\label{igualdad0}
  \overline{\bra \hat \pi_{\nk}^\RI \ket^2} = \overline{\bra (\hat \pi_{\nk}^\RI)^2 \ket} - \frac{1}{\textrm{Re} [A_k(\eta)]}.
\end{equation} 
The quantity $(\textrm{Re} [A(\eta)])^{-1} $ represents the standard
deviation of the squared momentum. It is also the width of every
packet in momentum space. The technical steps to obtain the right hand
side of eq.~\eqref{igualdad0} are presented in
ref.~\cite{CPS13}. However, in the present work we have generalized
those steps to the quasi-de Sitter case.

In particular, we need to use the expression for $A_k(\eta)$, eq.~\eqref{Ak}, and   find a suitable approximation  for the case  $-k\eta \ll 1$, i.e.
\begin{equation}\label{segundo}
  \frac{1}{\textrm{Re} [A_k(\eta)]} \simeq \frac{k 2^{2\mu-2} \sin(\pi \mu) \Gamma^2 (\mu) (-k\eta)^{-2\mu+1}   }{\pi \zeta_k^{2\mu} \sin (2\mu \theta_k + \pi \mu)},
\end{equation} 
where we have defined $\zeta_k e^{i \theta_k} \equiv \sqrt{1-2 i \lambda_k}$. Additionally, the quantity $ \overline{\bra (\hat \pi_{\nk}^\RI)^2 \ket} $ can be   approximated  for the case  $-k\eta \ll 1$, as
\begin{eqnarray}\label{primero}
  \overline{\bra (\hat \pi_{\nk}^\RI)^2 \ket}  &\simeq& \frac{k}{\pi} 2^{2\mu-2} \Gamma (\mu)^2 (-k\eta)^{-2\mu+1} \nn
  &\times& \bigg[ 1 + \lambda_k \sin \gamma_k \cos \gamma_k \nn
  &-& \frac{\lambda_k k \tau}{2} \bigg(  \frac{3}{\mu+1} \sin^2 \gamma_k + \frac{\cos^2 \gamma_k}{\mu}   \bigg) \bigg]
\end{eqnarray} 
with $\gamma_k \equiv -k \tau-\mu \pi/2 - 3\pi/4$. 

After inserting eqs.~\eqref{segundo} and \eqref{primero} into
eq.~\eqref{igualdad0}, one obtains the effective power spectrum from
eq.~\eqref{PS0}. The result is
\begin{equation}\label{PScslfinal}
   P(k) = A_s C(k) k^{n_s-1}.
\end{equation} 

As in the \emph{collapse schemes} approach, the amplitude predicted in
the CSL approach is the same as in the standard picture $A_s \propto
H^2/M_P^2 \epsilon_1$. Also, the prediction for the scalar spectral
index is different from the standard case, but identical to that
obtained for the \emph{collapse schemes} approach: $n_s-1 = 2
\epsilon_1 - \epsilon_2$. On the other hand, the function $C(k)$ reads
\begin{eqnarray}\label{Ck}
  C(k) &\equiv& 1 + \lambda_k k |\tau| + \lambda_k \cos(k |\tau| ) \sin (k |\tau| ) \nn
  &-& \frac{1}{\zeta_k^{2 n_s-1} \cos[ (2-n_s ) \theta_k]},
\end{eqnarray} 
where
\begin{equation}
  \zeta_k \equiv (1+4 \lambda_k^2 )^{1/4}, \qquad \theta_k \equiv -\frac{1}{2} \arctan (2\lambda_k).
\end{equation} 

It follows from the latter that when $\lambda_k =
\frac{\lambda_0}{k}$ 
the primordial power spectrum becomes nearly scale invariant, since
the most important dependence on $k$ arises from the second term. We
have checked that the changes resulting from the oscillatory terms of
eq.~\eqref{Ck} do not produce important effects in the
spectrum. Furthermore, it has been shown that, for the exact the
Sitter case the CMB spectrum is not sensitive to the value of
$\lambda_0$ \cite{LBL16}. We have also verified that this is the case
for the nearly invariant de Sitter case analyzed in this
paper. Furthermore, just as in the \emph{collapse schemes} approach,
one does not expect an exact $1/k$ dependence of $ \lambda_k $.  Thus
we proceed to explore possible effects on the shape of the primordial
power spectrum that would result from the following modified
dependence:

\begin{equation}
 \lambda_k \equiv \lambda_0 \left( \frac{1}{k} + \frac{\alpha}{k^2}   \right)
\end{equation} 
where we have introduced an extra parameter $\alpha$. The $\alpha/k^2$
term is motivated by the findings of a previous work \cite{BLA16}. In
such a work, a similar modified dependence was considered in the
context of the \emph{collapse schemes} approach.  The study indicated
a similar Ba\-ye\-sian evidence as the standard $\Lambda$CDM model.

\begin{figure}
\begin{center}
\includegraphics[scale=1.0]{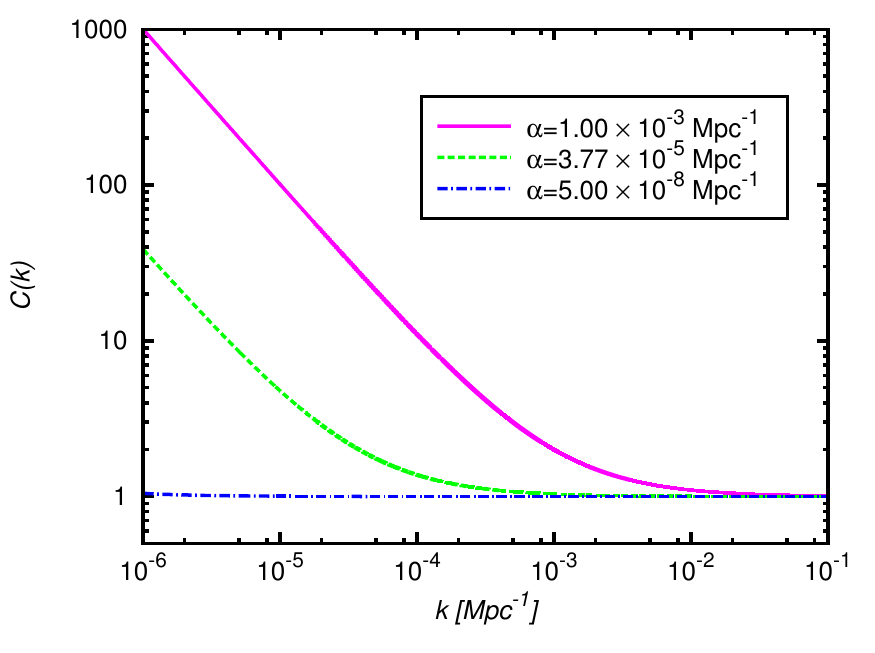}
\end{center}
\caption{The function $C(k)$ associated to the power spectrum in the
  CSL inflationary approach. We have set the value $\lambda_0 =
  1/|\tau| \simeq 6.41 \times 10^{-9}$ Mpc$^{-1}$ and $n_s = 0.96$. We
  have considered three different values of $\alpha$ (displayed in the
  figure). We have assumed standard values for $V_0$ and $N$.}
\label{figCkCSL}
\end{figure}

Figure \ref{figCkCSL}, shows the resulting function $C(k)$ for three
different values of $\alpha$. Note that $C(k) = 1$ means no
modification on the standard shape of the power spectrum. We consider
$\lambda_0=\frac{1}{\tau}$, where $\tau$ depends on two main
quantities: the characteristic energy scale of inflation $V_0$ and the
total number of e-folds of inflation $N$. We note that the effect of
considering the CSL model results in an important departure in both
the amplitude and shape of the large scale of the primordial power
spectrum, with the intensity depending on the value of $\alpha$. The
mentioned difference in shape between the standard primordial power
spectrum and the one resulting from the CSL collapse model is most
relevant for the lower values of $k$. On the other hand, $\lambda_k$
must be positive and this requirement implies $\alpha > -10^{-6}$ for
the relevant $k$ values.

\subsection{Effects of the self-induced collapse on the CMB spectrum}

Next, we explore the effects on the CMB spectrum, when incorporating
the self-induced collapse hypothesis within the various approaches
considered in this paper. Hereafter, we assume as a reference the
$\Lambda$CDM best-fit model reported by the Planck Collaboration
(2015) \cite{Ade:2015xua}.\footnote{We use the values obtained using
  the TT+lowP data.}  As regards the \emph{collapse schemes}, it has
already been shown in previous works \cite{LSS12,LLP15,BLA16} that if
\B$=0$ then the standard primordial power spectrum is recovered except
for an overall normalization factor, just as it does any change in the
collapse parameter \A.

\begin{figure}
\begin{center}
\includegraphics[scale=0.7]{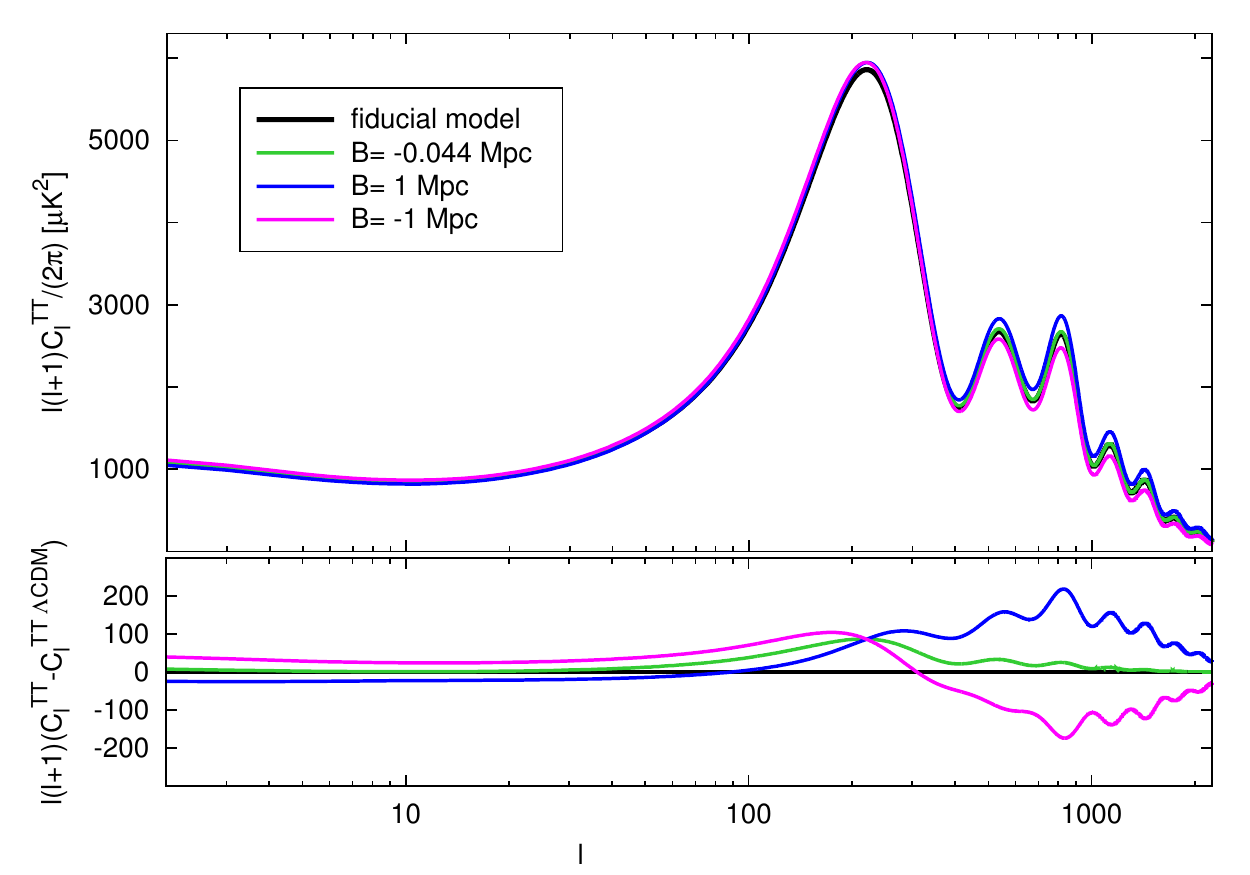}
\end{center}
\caption{The temperature auto-correlation function and differential plot respect to the fiducial   model for the \emph{Wigner} scheme model using \A$=-750$ and different values of \B. All
models are normalized to the maximum of the first peak of the fiducial model.}
\label{clswigner}
\end{figure}

\begin{figure}
\begin{center}
\includegraphics[scale=0.6]{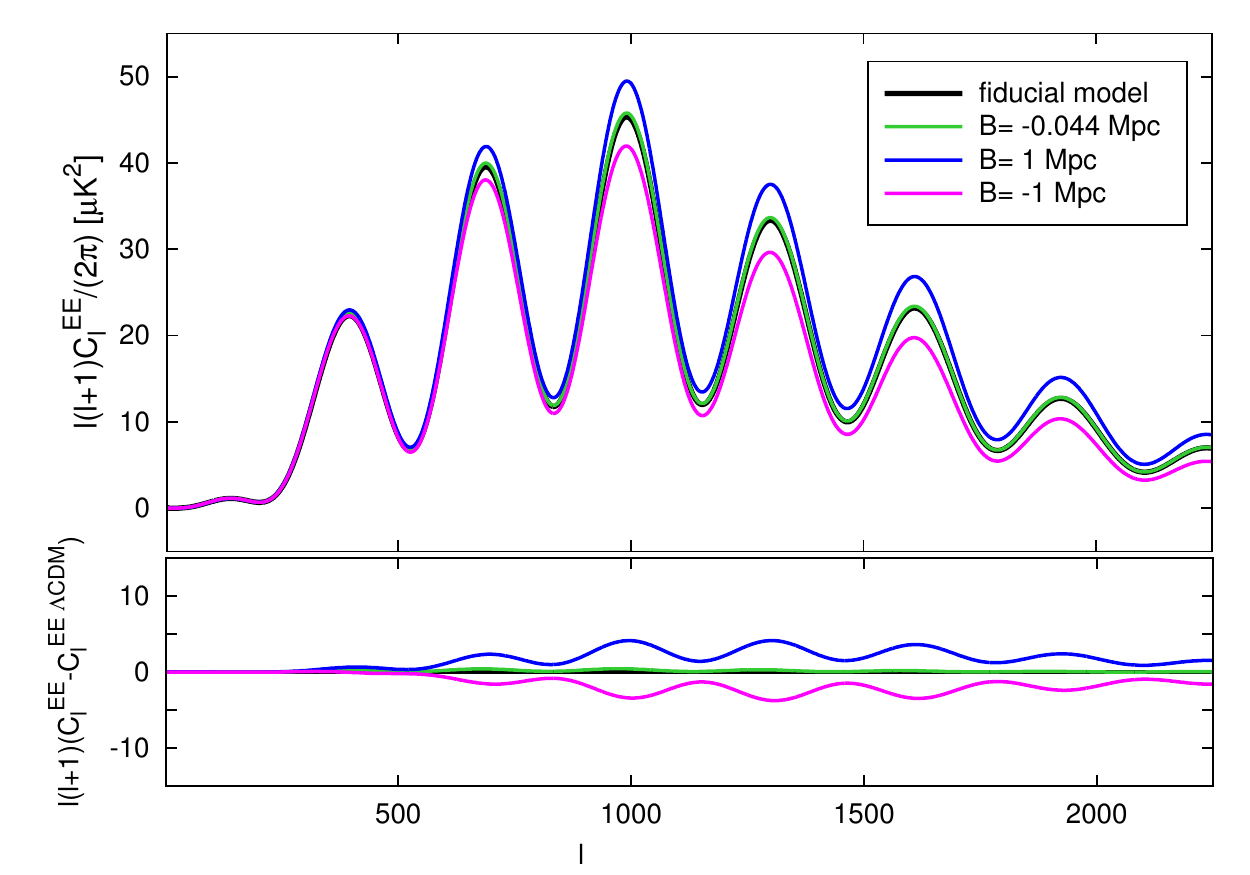}
\includegraphics[scale=0.6]{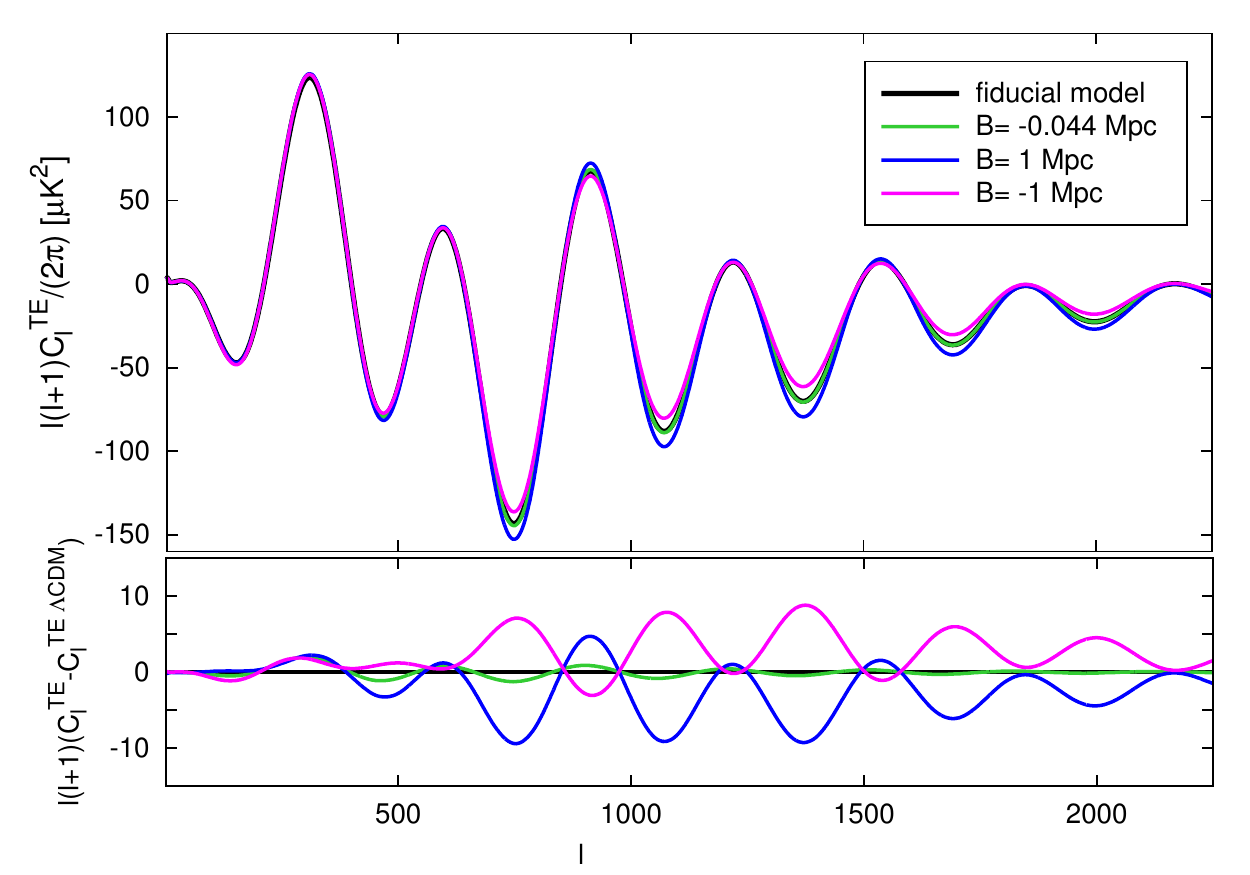}
\end{center}
\caption{Predictions for the \emph{Wigner} scheme model using
  \A$=-750$ and different values of \B. Left: The E-mode (EE)
  auto-correlation function and differential plot respect to the
  fiducial model f Right: The temperature-E mode polarization (TE)
  cross correlation power spectrum and differential plot respect to
  the fiducial model. All models are normalized to the maximum of the
  first peak of the fiducial model. }
\label{clswignerpol}
\end{figure}

Figures \ref{clswigner} and \ref{clswignerpol} show the CMB
temperature and polarization spectra for the \emph{Wigner} scheme
model using a fixed \A value and different values of \B. We choose to
fix \A to an ``appropriate" value (we stress that a change in \A
affects the {power spectrum just by an overall normalization}) so the
value of $A_s$ that gives a good fit to the CMB data is the closest to
the standard $\Lambda$CDM value.  At the same time, we chose the
values of the cosmological parameters for the collapse scheme models
to be the same as the ones of the fiducial model. We noted an increase
in the value of the secondary peaks of the temperature power spectrum
and a decrease in the values at the valleys for all cases, with the
magnitude of the changes depending on the value of \B. There is also
an increase in the height of the peaks in the EE spectrum, with
increasing values of \B while for the TE cross correlation temperature
we only noted a change in the height of the valleys with the the
intensity depending on the value of \B. A similar analysis was made
for the \emph{Newtonian} scheme model; the effects on the CMB spectrum
were similar, with the main difference being that in this scheme the
results are less sensitive to changes in \B \cite{LLP15}.

\begin{figure}
\begin{center}
\includegraphics[scale=0.67]{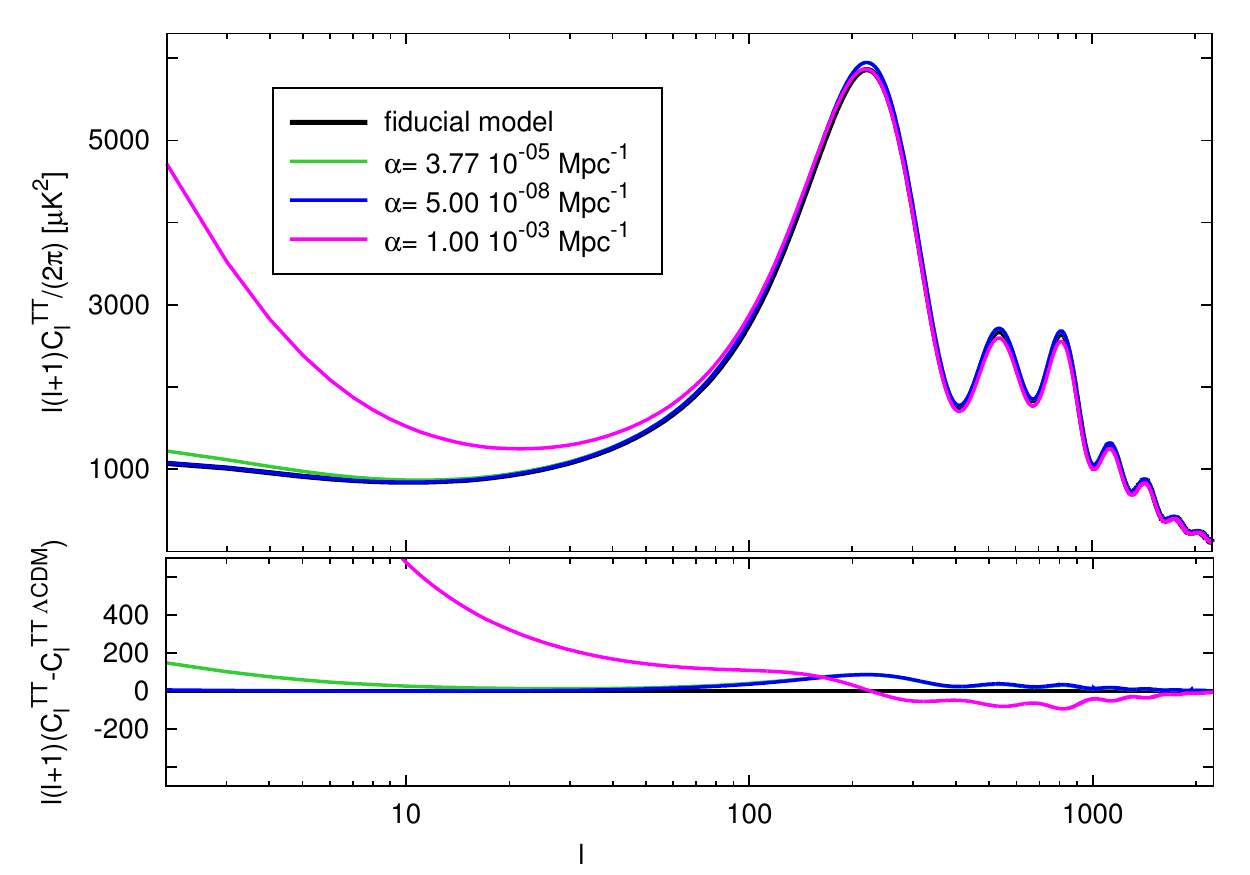}
\end{center}
\caption{The temperature (TT) auto-correlation power spectrum and
  differential plot respect to the fiducial model for the CSL model
  for different values of $\alpha$. All models are normalized to the
  maximum of the first peak of the fiducial model.}
\label{clscsl}
\end{figure}

\begin{figure}
\begin{center}
\includegraphics[scale=0.6]{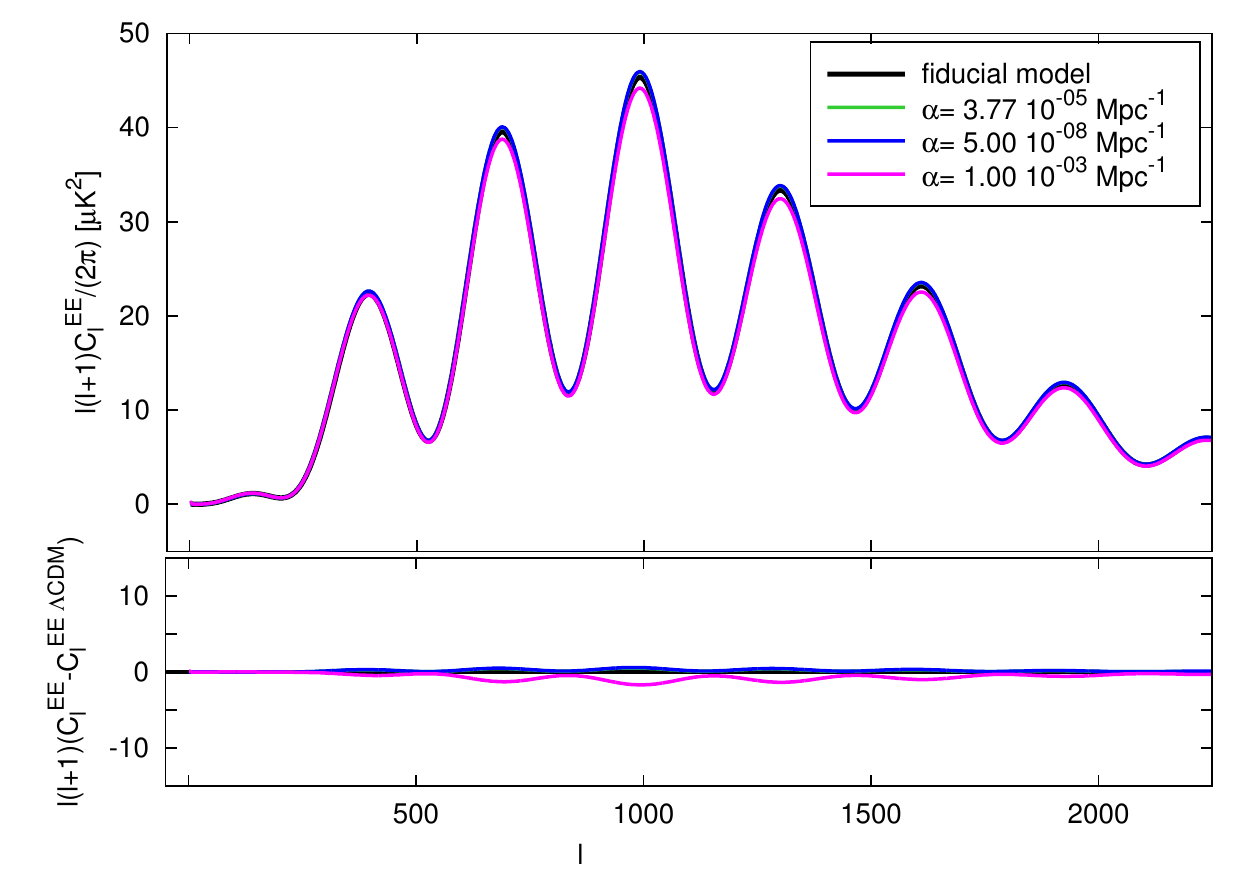}
\includegraphics[scale=0.6]{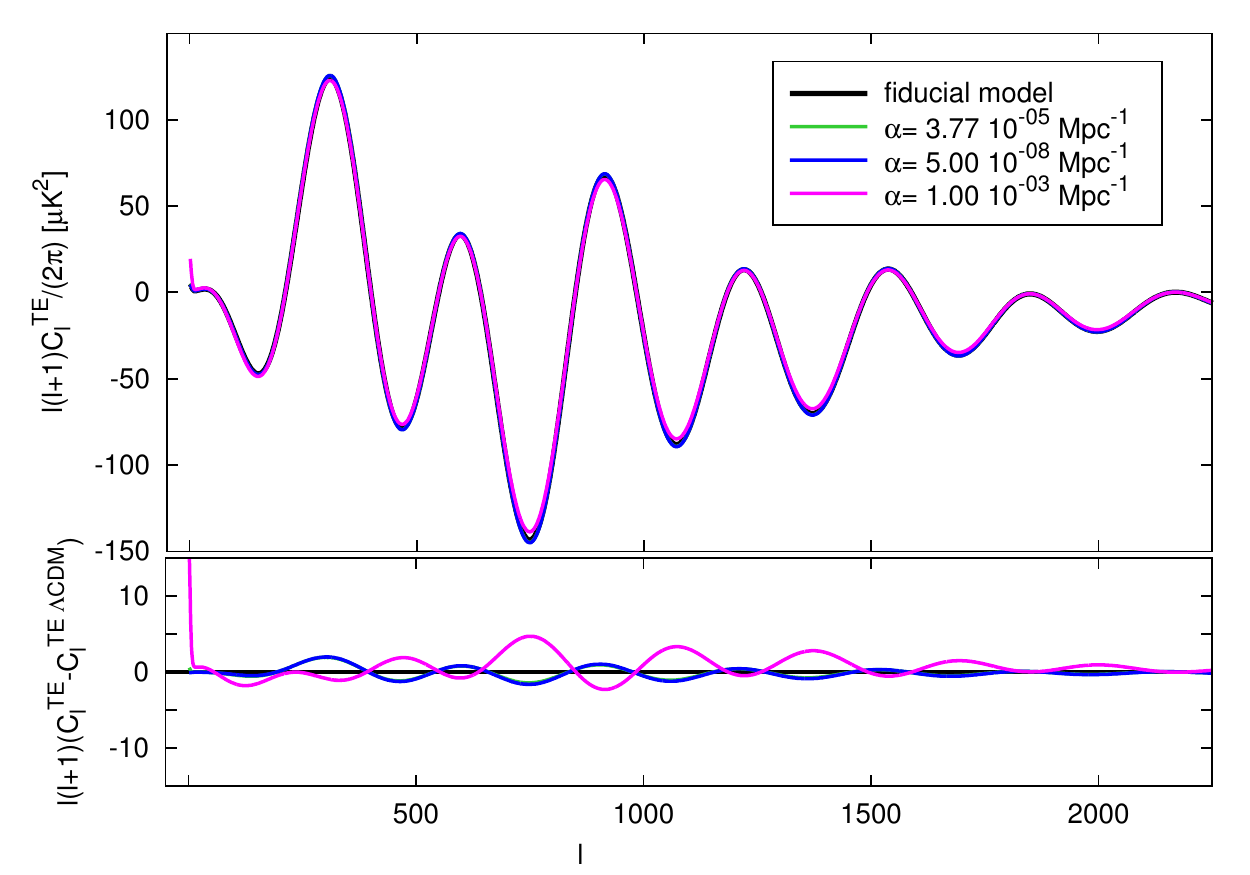}
\end{center}
\caption{Predictions for the CSL collapse model for different values
  of $\alpha$. Left: The E-mode polarization (EE) auto-correlation
  power spectrum and differential plot respect to the fiducial model
  Right: The temperature-E mode polarization (TE) cross correlation
  power spectrum and differential plot respect to the fiducial
  model. All models are normalized to the maximum of the first peak of
  the fiducial model.  }
\label{clscslpol}
\end{figure}

For the CSL collapse model, we note that a change in the parameter
$\alpha$ mainly affects the low multipole region as can be appreciated
in figure \ref{clscsl}.  However, we also observe a very small change
in the height of the peaks (with respect to the fiducial model) for
the E-mode spectrum while the temperature-polarization spectrum shows
no changes with respect to the fiducial model (see figure
\ref{clscslpol}). These effects are similar to the ones found in
\emph{collapse schemes} models with $\eta_c^k =
\frac{\mathcal{A}}{k}+\frac{\mathcal{B}}{k^2}$ (see
ref.~\cite{BLA16}).

\section{Analysis Method}\label{Method}

In this work we consider the two \emph{collapse schemes} models,
i.e. the \emph{Newtonian} and the \emph{Wigner} schemes, the CSL
collapse model and the $\Lambda$CDM one (that we use as a reference).

In our analysis, we vary the usual cosmological parameters, namely,
the physical baryon density, $\Omega_bh^2$, the physical cold dark
matter density, $\Omega_ch^2$, the ratio between the sound horizon and
the angular diameter distance at decoupling, $\theta$, the optical
depth, $\tau$, the primordial amplitude, $A_s$, the spectral index
$n_s$ and the additional collapse parameter \A, \B and $\alpha$.  We
also vary the nuisance foreground parameters~\cite{Aghanim:2015xee}
and consider purely adiabatic initial conditions. The sum of neutrino
masses is fixed to $0.06$ eV, and we limit the analysis to scalar
perturbations with $k_*=0.05$ $\rm{Mpc}^{-1}$.

We work with flat priors for the cosmological and collapse parameters,
and choose to fix the collapse parameter \A.  As has been discussed in
the previous section, the \A parameter just affects the primordial
spectrum as a change in the amplitude; therefore it is highly
degenerate with the $A_s$ parameter. Thus, we have tested several
values for the \A parameter, and fixed a value which satisfies the
condition for the conformal collapse time $\eta_c^k < 0$ and minimizes
the variation of the $A_s$ from the $\Lambda$CDM model value. (We
chose \A $= -750$ for the Wigner scheme and \A $= -600$ for the Newton
scheme.)

In order to compare the quantum collapse inflationary models with
recent CMB data, we need to compute CMB anisotropies including the
modifications in the primordial power spectrum. For this, we modify
the public available Code for Anisotropies in the Microwave Background
({\sc CAMB})~\cite{camb}.  In our analysis, we perform a Monte Carlo
Markov chain exploration of the parameters space using the available
package {\sc CosmoMC}~\cite{Lewis:2002ah} and implement the nested
sampling algorithm of {\sc Multinest}
code~\cite{Feroz:2008xx,Feroz:2007kg,Feroz:2013hea} to obtain the
Bayesian evidence of the model.  For the Bayesian analysis we use the
most accurate Importance Nested Sampling
(INS)~\cite{Cameron:2013sm,Feroz:2013hea} instead of the vanilla
Nested Sampling (NS), and we require a INS Global Log-Evidence error
$\leq 0.02$ .

For the data analysis, we use the current Planck Collaboration release
(2015)~\cite{Ade:2015xua} and BAO data. In particular, we consider the
high-$\ell$ Planck temperature data from the 100-,143-, and 217-GHz
half-mission T maps, and the low-$\ell$ data by the joint TT,EE,BB and
TE likelihood.  Also, we consider Baryonic Acoustic Oscillation data
by the 6dF Galaxy Survey (6dFGS)~\cite{Beutler:2011hx}, SDSS DR7 Main
Galaxy Sample (SDSS-MGS) galaxies~\cite{Ross:2014qpa}, BOSS galaxy
samples, LOWZ and CMASS~\cite{Anderson:2013zyy}.

We perform an appropriate comparison between the quantum collapse
inflationary models and the standard $\Lambda$CDM model predictions
using the Bayesian model comparison.  This is a powerful tool to favor
the models that fit well the data exhibiting strong predictivity,
while models with a large number of free parameters, not required by
the data, are penalised for the wasted parameter space (we refer the
reader to some recent employ on cosmological
models~\cite{Trotta:2005ar,Benetti:2016tvm,Benetti:2016ycg,Graef:2017cfy,Heavens:2017hkr,Campista:2017ovq}).

We can write the Bayes factor $\mathit{B}_{ij}=
\frac{\mathcal{E}_{M_i}}{\mathcal{E}_{M_j}}$, where
$\mathcal{E}_{M_i}$ is the evidence of the analysed model and
$\mathcal{E}_{M_j}$ the reference model one.  The usual scale employed
to judge the Bayes factor is the Jeffreys scale
\cite{Jeffreys,Trotta}, that is
\begin{center}
{\begin{tabular}{cccc}
$\ln{\mathcal{B}_{ij}}$ & Odds & Probability &Notes\\
\hline
$< 1$ 	&	$< 3 : 1$ & $< 0.750$ &\text{inconclusive}\\
\nonumber
$1$  & $\sim 3 : 1$ & $0.750$ &\text{weak evidence}\\
\nonumber
$2.5$ & $\sim 12 : 1$ & $ 0.923$ &\text{moderate evidence}\\
\nonumber
$5$    &$\sim 150 : 1$ &$0.993$ &\text{strong evidence}\\
\end{tabular}}
\end{center}
Note that negative Bayes factor value means support in favor of the
reference model $j$.

\section{Results}
\label{results}

\begin{figure}
\begin{center}
\includegraphics[scale=0.8]{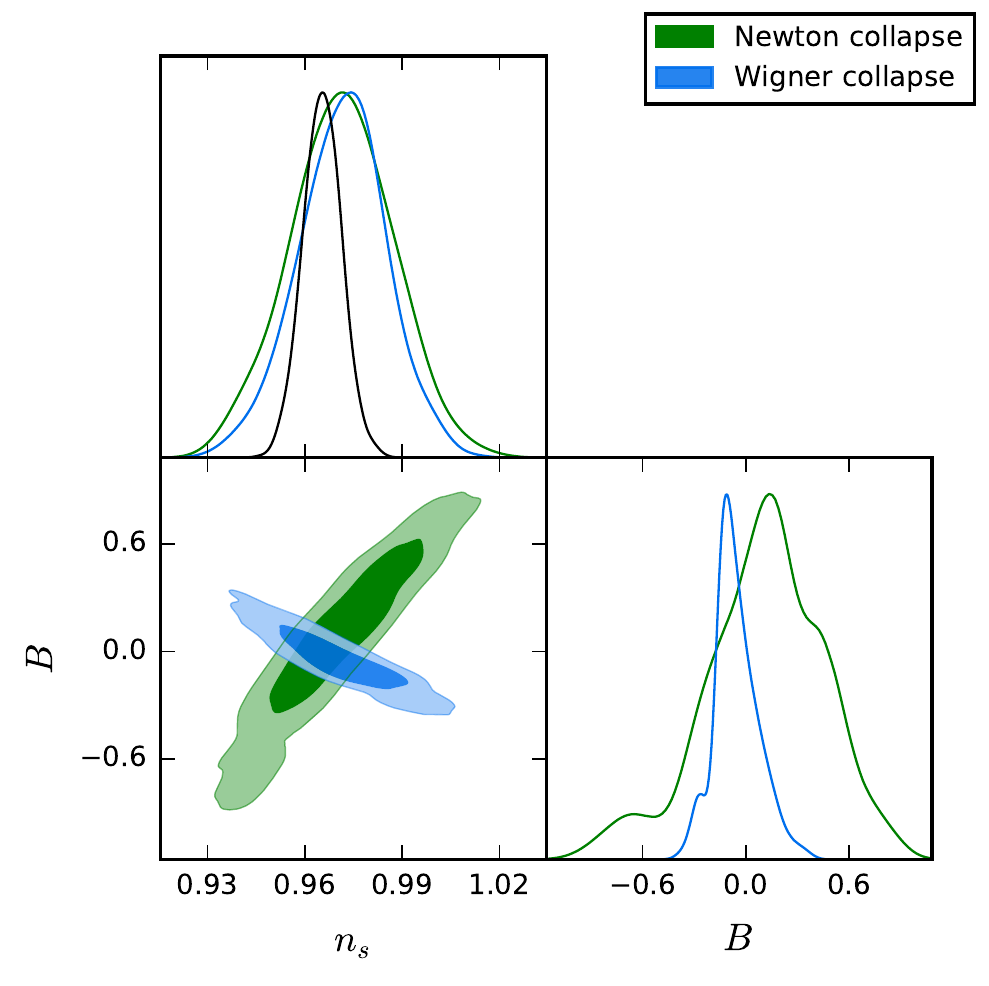}
\end{center}
\caption{$68\%$ and $95\%$ confidence regions in the $n_s-B$ plane for the \emph{Newtonian} (green curve) and \emph{Wigner} (blue curve) collapse model. Results using the CMB data. The black curve refers to the $\Lambda$CDM model.}
\label{fig:tri_new_wig}
\end{figure}

Before presenting our results, let us draw attention on the different
degeneracy between $n_s$ and \B for the two \emph{collapse schemes}
models analysed. In figure \ref{fig:tri_new_wig}, we can observe that,
for the \emph{Newtonian} scheme (green curve), increasing values of
$n_s$ allow for higher values of \B; while in the \emph{Wigner} scheme
(blue curve) for crescent values of $n_s$, lower values of \B are
preferred by the data.  Furthermore, the \emph{Newtonian} scheme
allows for high values of the parameter $n_s$ (until the unity) and it
is interesting in the context of the so-called $H_0$ tension. Indeed,
the degeneracy between the spectral index parameter and the local
value of the Hubble constant, (i.e. higher value of $n_s$ produces an
increase in the value of $H_0$), reduces the tension between the $H_0$
value derived by CMB analysis and the local measurements of
\emph{Riess et al.} from the Hubble Space Telescole
(HST)~\cite{Riess09} (see
refs.~\cite{Bernal:2016gxb,Benetti:2017gvm,Zhang:2017epd,Addison:2017fdm}
for recent discussions about the $H_0$ tension).

We present the cosmological analysis for the \emph{collapse schemes}
models in tables \ref{tab:results_cmb1} and \ref{tab:results_cmb2}. We note that the resulting
constraints on the parameters' values are in general agreement with
those obtained for the $\Lambda$CDM model. However, it should be noted
that $n_s$ and $A_s$ are less constrained than in the standard model
due to the degeneration of these parameters with \B.  On the other
hand, while the main value of the primordial amplitude is in agreement
with that obtained for the $\Lambda$CDM model in the \emph{Wigner}
model, we note a shift in the one obtained for the \emph{Newtonian}
scheme.  Furthermore, we also note that the constraints on \B are
narrower in the \emph{Wig\-ner} scheme than the \emph{Newtonian}
one. This reflects the increased sensibility of the observational
predictions of this model to small variations of the \B parameter
value.

In the last lines of tables \ref{tab:results_cmb1} and \ref{tab:results_cmb2} we report the
$\Delta\chi^2$ and the Bayes factor $\ln{\mathcal{B}_{ij}}$ for the
models with respect to the standard cosmological one.  For the
\emph{Newtonian} scheme, the $\chi^2$ value is better than the
$\Lambda$CDM one of one point while in the \emph{Wigner} scheme the
improvement over the $\Lambda$CDM model is $1.9$. However, the data
show a \emph{moderate} preference for the $\Lambda$CDM model over the
\emph{Newtonian} model and a \emph{weak} preference over the
\emph{Wigner} one.  This is due to the spread in the non-gaussian
profile of the posterior probability distributions for \B.

\begin{figure}
\begin{center}
\includegraphics[scale=0.6]{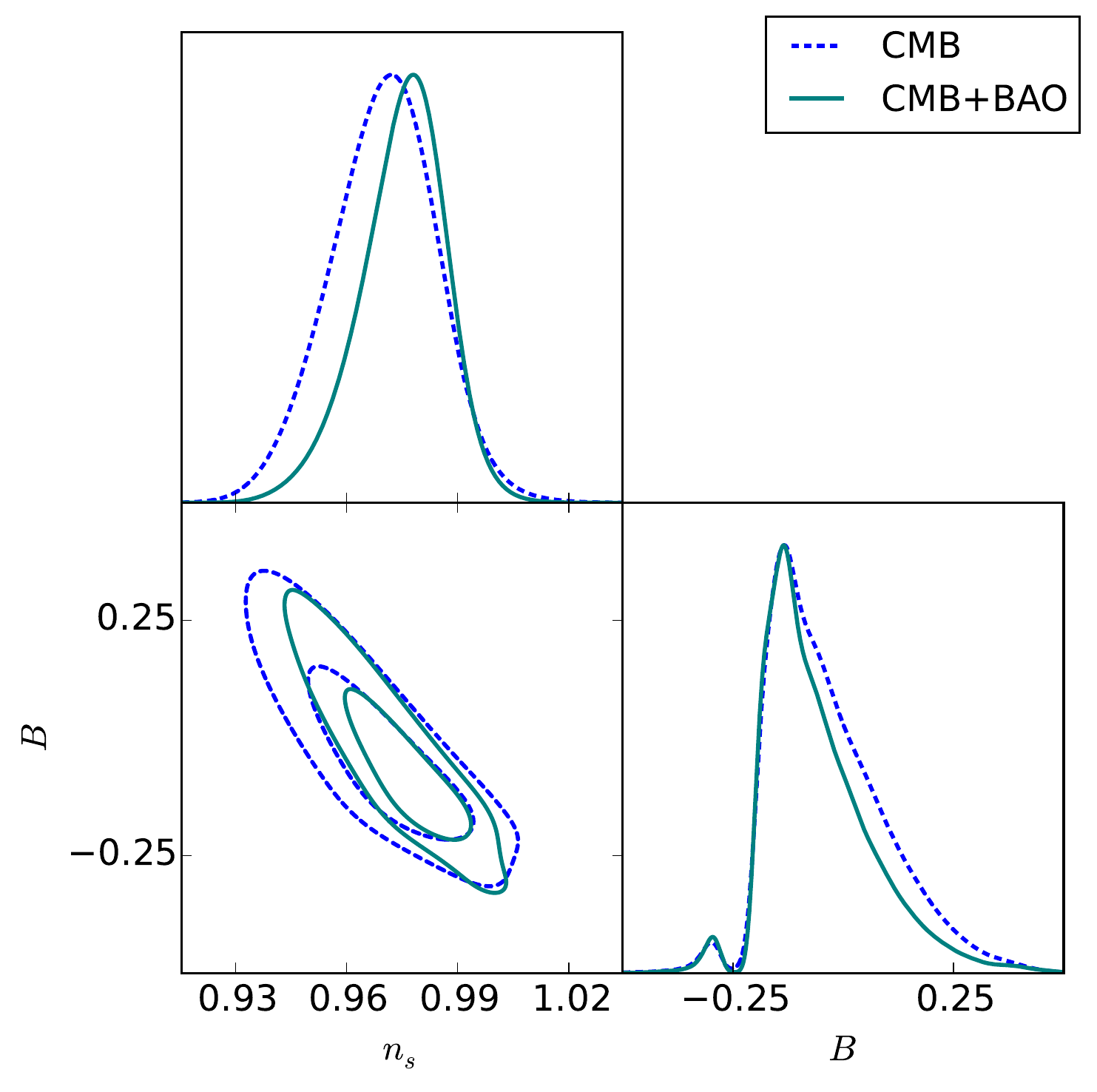}
\end{center}
\caption{$68\%$ and $95\%$ confidence regions in the $n_s-B$ plane for the \emph{Wigner} collapse model . One dimensional probability densities for the $B$  and $n_s$ parameter of the CSL collapse model. Results using only CMB and CMB+BAO data are shown.}
\label{fig:tri_wigner_cmb_bao}
\end{figure}

In order to improve these results, we also compared the predictions of
the \emph{collapse scheme} model with the BAO data. Indeed, the
imprint of baryon acoustic oscillations in large-scale structure are a
powerful tool for mapping out the cosmic expansion history and
constrain the cosmological parameters.  Given that the \emph{Wig\-ner}
scheme is more sensitive to changes with the \B parameter, and also
shows a better Bayesian evidence than the \emph{Newtonian} scheme, we
select it for this second analysis. In table \ref{tab:results_cmb_bao}
we report the results for the joint CMB and BAO data set. We can see
that the values of the cosmological parameters are more strongly
constrained
but, at the same time the constraints on \B show almost no difference
when the BAO data are considered (see also the figure
\ref{fig:tri_wigner_cmb_bao}). Furthermore, the data show
\emph{mo\-de\-rate} Bayesian preference for the \emph{Wig\-ner} scheme
model over the $\Lambda$CDM model for the CMB+BAO data.  We stress
that, comparing with the Bayesian evidence of the tables
\ref{tab:results_cmb1} and \ref{tab:results_cmb2}, the \emph{Wigner} scheme shows very closed
value while the $\Lambda$CDM model gets a value worst of about 6
points. This means that the improving in the $B_{ij}$ is mainly due to
a worst fit of the $\Lambda$CDM data of the new dataset, while the
\emph{Wigner} scheme proves to be more conservative.

\begin{figure}
\begin{center}
\includegraphics[scale=0.6]{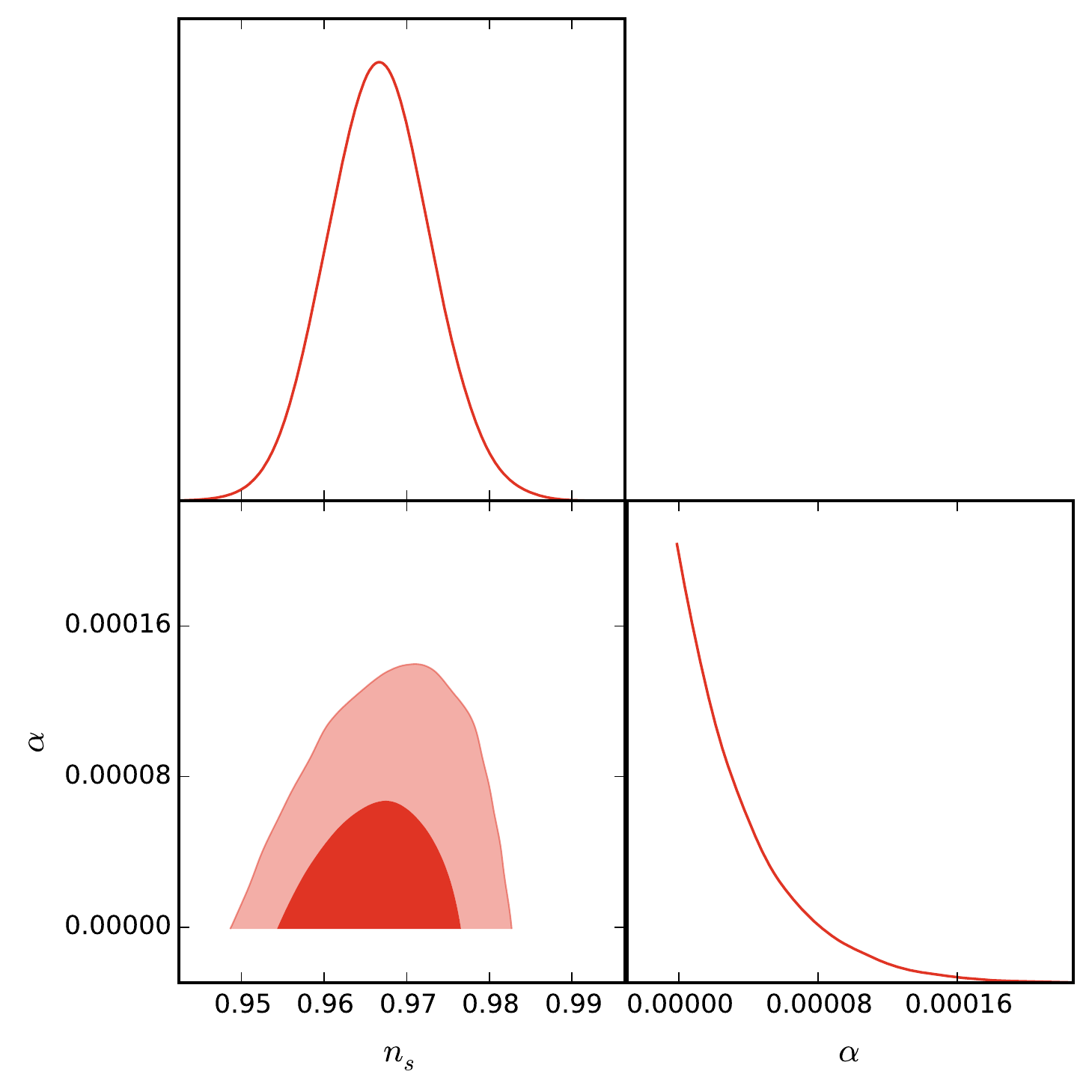}
\end{center}
\caption{$68\%$ and $95\%$ confidence regions in the $n_s-B$ plane for the CSL collapse model . One dimensional probability densities for the $B$  and $n_s$ parameter of the CSL collapse model.}
\label{fig:tri_csl}
\end{figure}

Now, we look to the cosmological analysis of the CSL collapse
model. The results are in table \ref{tab:results_csl}, where we note
an excellent agreement between the parameters values obtained from the
analysis of the CSL collapse and the $\Lambda$CDM model.  Moreover, in
this case there is no degeneration between $n_s$ and the $\alpha$
collapse parameter (see figure \ref{fig:tri_csl}) and no increase of
$n_s$ constraints is encountered for this model. On the other hand,
the $\alpha$ parameter is not well constrained, and just an upper
bound was obtained from the statistical analysis.  In this case, we do
not analyze the CSL model using BAO data since the theoretical
prediction of this model differs from the standard model one only at
very large angular scales. On the other hand, it is well known that
BAO bumps are observed only at low angular scales, which means that
the BAO data provide no useful information from large angular scales
\cite{Carvalho2016}.  Finally, we note that the data show the same
Bayesian preference for this model and the standard cosmological one.

\begin{table*}[h]
  \caption{$68\%$ confidence limits for the cosmological and collapse scheme parameters. The first columns-block refer to the minimal $\Lambda$CDM model; 
    the second block shows the constraints on the \emph{Newtonian} models; 
    $\Delta \chi^2_{best} = \chi^2_{\rm best}(\Lambda CDM) - \chi^2_{\rm best}({\rm collapse \,\,\, model})$; For  $\ln {B}_{ij}$, the reference model is $\Lambda$CDM.} 
{\begin{tabular*}{\textwidth}{@{\extracolsep{\fill}}lrrrr@{}}\hline
\multicolumn{1}{c}{$ $}&
\multicolumn{2}{c}{\textbf{$\Lambda$CDM model}}& 
\multicolumn{2}{c}{\textbf{\emph{Newtonian}-scheme}}
\\ 
Parameter & mean & bestfit & mean & bestfit \\
\hline
$100\,\Omega_b h^2$ 	
& $2.223 \pm 0.023$ & $2.218 $ 		
& $2.231 \pm 0.035$ & $2.239$     
\\
$\Omega_{c} h^2$	
& $0.1197 \pm 0.0022$ & $0.1199 $	
& $0.1194 \pm 0.0023$ & $0.1178$        
\\
$100\, \theta$ 
& $1.04087 \pm 0.00048$ & $1.04070 $	
& $1.04094 \pm 0.00050$ & $1.04118 $    
\\
$\tau$
& $0.078 \pm 0.020$& $0.082$	
& $0.078 \pm 0.021$& $0.089$    
\\
$n_s$ 
& $0.9656 \pm 0.0064$ & $0.9640 $	
& $0.9670 \pm 0.0147$ & $0.9802$ 
\\
$\ln 10^{10}A_s$  \footnotemark[1]
\footnotetext[1]{$k_0 = 0.05\,\Mpc^{-1}$.}
& $3.091 \pm 0.037$ & $ 3.100 $	
& $4.108 \pm 0.042$ & $4.128$   
\\
 \B 	 
& $-$ & $-$
& $0.050 ^{+0.350}_{-0.311}$& $0.206$
\\
\hline
\hline
$\Delta \chi^2_{\rm best}$         
& & $-$	%
& & $1$ 
\\
$\ln \mathit{B}_{ij}$ 
& & $-$ 
& &$-3.33 $
\\
\hline
\end{tabular*} \label{tab:results_cmb1}}
\end{table*} 

\begin{table*}[h]
  \caption{$68\%$ confidence limits for the cosmological and collapse scheme parameters. The first columns-block refer to the minimal $\Lambda$CDM model; 
    the second block shows the constraint on the  \emph{Wigner} scheme models; 
    $\Delta \chi^2_{best} = \chi^2_{\rm best}(\Lambda CDM) - \chi^2_{\rm best}({\rm collapse \,\,\, model})$; For  $\ln {B}_{ij}$, the reference model is $\Lambda$CDM.} 
{\begin{tabular*}{\textwidth}{@{\extracolsep{\fill}}lrrrr@{}}
\hline
\multicolumn{1}{c}{$ $}&
\multicolumn{2}{c}{\textbf{$\Lambda$CDM model}}& 
\multicolumn{2}{c}{\textbf{\emph{Wigner}-scheme}}
\\ 
Parameter	& mean & bestfit & mean & bestfit  \\
\hline
$100\,\Omega_b h^2$ 	
& $2.223 \pm 0.023$ & $2.218 $ 		
& $2.231 \pm 0.031$ & $2.253 $          
\\
$\Omega_{c} h^2$	
& $0.1197 \pm 0.0022$ & $0.1199 $	
& $0.1194 \pm 0.0023$ & $0.1192 $       
\\
$100\, \theta$ 
& $1.04087 \pm 0.00048$ & $1.04070 $	
& $1.04093 \pm 0.00049$ & $1.04105 $ 
\\
$\tau$
& $0.078 \pm 0.020$& $0.082$	
& $0.080 \pm 0.021$& $0.077$    
\\
$n_s$ 
& $0.9656 \pm 0.0064$ & $0.9640 $	
& $0.9700 \pm 0.0157$ & $0.9778 $ 
\\
$\ln 10^{10}A_s$  \footnotemark[1]
\footnotetext[1]{$k_0 = 0.05\,\Mpc^{-1}$.}
& $3.091 \pm 0.037$ & $ 3.100 $	
& $3.065 \pm 0.056$ & $ 3.080$ 	
\\
 \B 	 
& $-$ & $-$
& $-0.037 ^{+0.084}_{-0.160} $ & $-0.139$ 
\\
\hline
\hline
$\Delta \chi^2_{\rm best}$         
& & $-$	%
& & $1.9$ 
\\
$\ln \mathit{B}_{ij}$ 
& & $-$ 
& &$-1.44 $ 
\\
\hline
\end{tabular*} \label{tab:results_cmb2}}
\end{table*}

\begin{table*}
  \caption{
    $68\%$ confidence limits for the cosmological and collapse scheme parameters using  CMB and BAO data. 
    The first columns-block refer to the minimal $\Lambda$CDM model; the second block shows the constraint on the  \emph{Wigner} scheme models; 
    $\Delta \chi^2_{best} = \chi^2_{\rm best}(\Lambda CDM) - \chi^2_{\rm best}({\rm collapse \,\,\, model})$; For  $\ln {B}_{ij}$, the reference model is $\Lambda$CDM.} 
{\begin{tabular*}{\textwidth}{@{\extracolsep{\fill}}lrrrr@{}}
\hline
\multicolumn{1}{c}{$ $}&
\multicolumn{2}{c}{\textbf{$\Lambda$CDM model}}&
\multicolumn{2}{c}{\textbf{\emph{Wigner}-scheme}}
\\
Parameter    & mean & bestfit & mean & bestfit  \\
\hline
$100\,\Omega_b h^2$   
& $2.233 \pm 0.020$ & $2.234 $         
& $2.243 \pm 0.025$ & $2.256 $          
\\
$\Omega_{c} h^2$  
& $0.1181 \pm 0.0012$ & $0.1175 $    
& $0.1180 \pm 0.0012$ & $0.1180 $       
\\
$100\, \theta$
& $1.04111 \pm 0.00041$ & $1.04089 $    
& $1.04112 \pm 0.00042$ & $1.04106 $ 
\\
$\tau$
& $0.084 \pm 0.018$& $0.091$    
& $0.086 \pm 0.019$& $0.102$    
\\
$n_s$
& $0.9696 \pm 0.0043$ & $0.9692 $    
& $0.9752 ^{+0.0132}_{-0.0090} $ & $0.9819 $ 
\\
$\ln 10^{10}A_s$  \footnotemark[1]
\footnotetext[1]{$k_0 = 0.05\,\Mpc^{-1}$.}
& $3.099 \pm 0.035$ & $ 3.110 $ 
& $3.082 \pm 0.050$ & $ 3.127$  
\\
 \B     
& $-$ & $-$
& $-0.058 ^{+0.071}_{-0.144} $ & $-0.134$ 
\\
\hline
\hline
$\Delta \chi^2_{\rm best}$         
& $-$ & 	%
&  & $0.3$
\\
$\ln \mathit{B}_{ij}$ 
& & $-$ 
& & $2.96$
\\
\hline
\end{tabular*}\label{tab:results_cmb_bao}}
\end{table*}

\begin{table*}
\caption{
$68\%$ confidence limits for the cosmological and CSL collapse model  using  CMB  data. 
The first columns-block refer to the minimal $\Lambda$CDM model; the second block shows the constraint on the  CSL collapse model; 
$\Delta \chi^2_{best} = \chi^2_{\rm best}(\Lambda CDM) - \chi^2_{\rm best}({\rm collapse \,\,\, model})$; For  $\ln {B}_{ij}$, the reference model is $\Lambda$CDM. }
{\begin{tabular*}{\textwidth}{@{\extracolsep{\fill}}lrrrr@{}}\hline
\multicolumn{1}{c}{$ $}&
\multicolumn{2}{c}{\textbf{$\Lambda$CDM model}}&
\multicolumn{2}{c}{\textbf{\emph{CSL} model}}
\\
Parameter    & mean & bestfit & mean & bestfit   \\
\hline
$100\,\Omega_b h^2$   
& $2.223 \pm 0.023$ & $2.218 $    
& $2.222 \pm 0.023$ & $2.234$     
\\
$\Omega_{c} h^2$  
& $0.1197 \pm 0.0022$ & $0.1199 $    
& $0.1197 \pm 0.0022$ & $0.1192 $    
\\
$100\, \theta$
& $1.04087 \pm 0.00048$ & $1.04070 $    
& $1.04087 \pm 0.00048$       & $1.04064 $    
\\
$\tau$
& $0.078 \pm 0.020$ & $0.082$    
& $0.075 \pm 0.019$      & $0.087$    
\\
$n_s$
& $0.9656 \pm 0.0064 $ & $0.9640 $    
& $0.9667 \pm 0.0062$      & $0.9674$     
\\
$\ln 10^{10}A_s$  \footnotemark[1]
\footnotetext[1]{$k_0 = 0.05\,\Mpc^{-1}$.}
& $3.091 \pm 0.037$ & $ 3.100 $ 
& $3.082 \pm 0.037$ & $ 3.112$ 
\\
$10^5 \alpha$     
& $-$ & $-$
& $ < 4.3 $ & $0.95$ 
\\
\hline
\hline
$\Delta \chi^2_{\rm best}$         
& & 
& & $-0.622$
\\
$\ln \mathit{B}_{ij}$ 
& &  
& & $ -0.3 $
\\
\hline
\end{tabular*}\label{tab:results_csl}}
\end{table*}

%
%

\section{Conclusions}

\label{conclusions}

In this paper, we have studied the phenomenological predictions of two
different collapse proposals: the \emph{collapse scheme} approach and
the Continuous Spontaneous Localization inflationary collapse
approach. For the former, we have considered the \emph{New\-to\-nian}
and \emph{Wig\-ner} collapse schemes with the collapse time
$\tc={\mathcal A}/{k}+\mathcal B$. For the latter, we have considered
the conjugated momentum of the inflaton field as the collapse
operator.

We have performed a statistical analysis in order to compare the
predictions of the theoretical models with recent CMB and BAO
data. Our findings indicate that collapse inflationary models are
compatible, for the appropriate choice of the values of the free
parameters with recent CMB and BAO data.  Furthermore, for the
collapse schemes considered in this work we have obtained stringent
bounds on the collapse parameter \B which characterizes the dynamics
of collapse time of each mode $\tc$ .  We have also obtained an upper
bound on the parameter $\alpha$ of the CSL model which is related to
the strength of the collapse. In addition, the values obtained for the
cosmological parameters are consistent with those obtained by the
Planck collaboration assuming a standard inflationary scenario.  On
the other hand, the constraints obtained for $n_s$, within the
\emph{collapse schemes} approach, are less stringent than those
obtained in the context of the standard inflationary scenario. As a
consequence, inflationary potentials that were discarded in such a
context, could be reconsidered in the collapse proposal. Finally,
results from the Bayesian model comparison method show a preference of
the \emph{Wigner} collapse model over the reference model when BAO
data are included in the analysis, while there is no such preference
when just the CMB data are considered. Moreover, the CSL collapse
model gives the same Bayesian evidence as the standard $\Lambda$CDM
model, while the latter is preferred over the \emph{Newtonian} scheme
collapse model.

%
\section*{Acknowledgements}

The authors acknowledge the use of the supercluster Mitzli at UNAM for
the statistical analyses and thank the people of DGSCA-UNAM for
computational and technical support.  MB acknowledges financial
support from the Funda\c{c}\~{a}o Carlos Chagas Filho de Amparo \`{a}
Pesquisa do Estado do Rio de Janeiro (FAPERJ - fellowship \emph{Nota
10}), and is also supported by INFN, Naples section, QGSKY project.
MPP, GL and SL are supported by PIP 11220120100504 CONICET and by the
National Agency for the Promotion of Science and Technology (ANPCYT)
of Argentina grant PICT-2016-0081; and by grant G140 from UNLP.  DS
acknowledges partial financial support from DGAPA-UNAM project
IG100316 and by CONACYT project 101712, the FAE-Network of CONACYT, as
well as the sabbatical fellowships from CO-MEX-US (Fullbright--Garcia
Robles) and from DGAPA-UNAM (Paspa).  The authors acknowledge the use
of CosmoMC~\cite{Lewis:2002ah} and Multinest
code~\cite{Feroz:2008xx,Feroz:2007kg,Feroz:2013hea}.


\bibliography{bibliografia}   

\end{document}